\documentclass[12pt]{iopart}

\usepackage{graphicx}

\eqnobysec

\begin{document}

\title[Adsorption of self-interacting polymers on fractals]
{Exact and Monte Carlo study of adsorption  of a self-interacting
polymer chain for a family of three-dimensional fractals}

\author{S Elezovi\'c-Had\v zi\'c\dag, I \v Zivi\'c \ddag  and S Milo\v sevi\'c\dag}

\address{\dag Faculty of Physics,
University of Belgrade, P.O.Box 368, 11001 Belgrade, Serbia,
Yugoslavia} \address{ \ddag Faculty of Natural Sciences and
Mathematics, University of Kragujevac, 34000 Kragujevac, Serbia,
Yugoslavia}

\eads{\mailto{sunchica@net.yu}, \mailto{ivanz@knez.uis.kg.ac.yu},
\mailto{emilosev@etf.bg.ac.yu}}

\begin{abstract}
We study the problem of adsorption of self-interacting
linear polymers situated in fractal containers that belong
to the three-dimensional (3d) Sierpinski gasket (SG) family
of fractals. Each member of the 3d SG fractal family has a
fractal impenetrable 2d adsorbing surface (which is, in
fact, 2d SG fractal) and can be labelled by an integer $b$
($2\le b\le\infty$). By applying the exact and Monte Carlo
renormalization group (MCRG) method, we calculate the
critical exponents $\nu$ (associated with the mean squared
end-to-end distance of polymers) and $\phi$ (associated
with the number of adsorbed monomers), for a sequence of
fractals with $2\le b\le4$ (exactly) and $2\le b\le40$
(Monte Carlo). We find that both $\nu$ and $\phi$
monotonically decrease with increasing $b$ (that is, with
increasing of the container fractal dimension $d_f$), and
the interesting fact that both functions, $\nu(b)$ and
$\phi(b)$, cross the estimated Euclidean values. Besides,
we establish the phase diagrams, for fractals with $b=3$
and $b=4$, which reveal existence of six different phases
that merge together at a multi-critical point, whose nature
depends on the value of the monomer energy in the layer
adjacent to the adsorbing surface.

\end{abstract}
\pacs{ 64.60.-i, 36.20.-r, 05.50.+q}
\submitto{\JPA}

\maketitle

\section{Introduction}
\label{uvod}

Statistics of a polymer chain in various types of solvents near an
impenetrable wall (boundary) with short--range attractive forces
has been extensively studied because of its practical importance
\cite{napper}, and as a challenging problem within the modern
theory of critical phenomena \cite{binder}. The most frequently
applied model for a polymer chain has been the self--avoiding
(SAW) random walk model (that is, the walk without
self--intersections), so that steps of the walk have been
identified with monomers that comprise the polymer, while the
solvent surrounding has been represented by a lattice. These
problems have been mostly studied for models situated on
two--dimensional (2d) Euclidean lattices using various theoretical
methods, such as the series expansion method, the renormalization
group (RG) techniques, the mean--field approach, Monte Carlo
simulations, and the conformal invariance method. In the last two
decades these problems have been also studied in a case of fractal
lattices embedded in the two--dimensional Euclidean space. On the
other hand, in a more realistic three--dimensional case (for the
Euclidean lattices, as well as for fractal lattices), a smaller
number of results have been obtained. The study of fractal
lattices has an advantage not only because their intrinsic
self--similarity makes the problem more amenable to an exact
approach, but also because these lattices as such may serve to
model porous media.

In this paper we report results of our study of a linear
polymer situated in the three-dimensional (3d) fractal
lattices that belong to the Sierpinski gasket (SG) family
of fractals, assuming the interaction between two adjacent
nonconsecutive monomers and, in addition, assuming
adsorbing interaction with the walls of the fractal
interior. This problem has been studied in the case of the
3d Euclidean lattices (see, for instance  \cite{Dhar2001},
and references quoted therein), but the number of obtained
results is definitely smaller than in the corresponding 2d
case. In the 3d case, the main endeavor has been manifested
in attempts to establish phase diagrams
\cite{Dhar2001,Vrbova} in the interaction parameter space
(which consists of the monomer--monomer interaction
parameter and the adsorption energy parameter). In addition
to the phase diagram, attempts have been made to calculate
critical exponents that characterize various polymer
configurations. The first work of this kind for a fractal
was done by Bouchaud and Vannimenus \cite{Bouchaud}, who
applied the renormalization group (RG) technique for the 3d
SG with scale parameter $b=2$. Here we report results of
our exact RG calculation for the 3d SG with $b=3$ and
$b=4$, and our results for the sequence $2\le b\le40$
obtained via the Monte Carlo renormalization group (MCRG)
method. Therefore, this paper appears to reflect an effort
to extend our previous studies performed  in the case of
the two--dimensional fractal lattices. Such an effort has
been incited by the well--known fact that critical
properties of a given model depend on dimension of the
space in which the model is situated.

This paper is organized as follows. In Sec.~\ref{druga} we
first describe the 3d SG fractals for general $b$. Then, we
present the framework of the RG method for studying the
polymer adsorption problem on these fractals (taking into
account the presence of the monomer--monomer interaction),
in a way that should make the method transparent for exact
calculations, as well as for the Monte Carlo calculations.
In Sec.~\ref{treca} we elaborate on the phase diagrams
obtained through the exact RG analysis for the $b=3$ and
$b=4$ SG fractals, and, in addition, we display our
findings for the concomitant critical exponents $\nu$
(associated with the mean squared end-to-end distances of
polymers) and $\phi$ (associated with the number of
adsorbed monomers). For $b=3$, values of $\nu$, for
different phases, have been previously calculated
\cite{Knezevic}, so that here we report the values for the
corresponding crossover exponent $\phi$, whereas in the
$b=4$ case we had to calculate values for the both
exponents $\nu$ and $\phi$. In Sec.~\ref{mcrg} we explain
details of the MCRG calculations of the critical exponents,
for arbitrary $b$, and present their specific values up to
$b=40$. Summary of the obtained results and the relevant
conclusions are given in Sec.~\ref{diskusija}.

\section{ Renormalization group scheme}
\label{druga}

\begin{figure}
\hskip4cm
\includegraphics[scale=0.3]{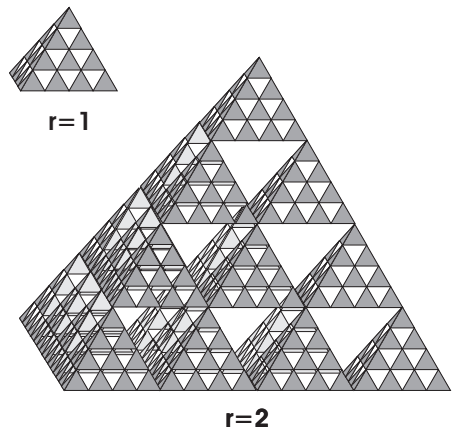}
\caption{The first two steps ($r=1$ and $r=2$) of the
self--similar construction of the 3d SG fractals, in the case
$b=4$.} \label{fig1}
\end{figure}

Each member of the 3d SG family of fractals is labeled by
the scale parameter $b=2,3,4,\ldots$ and can be constructed
recursively starting with the pertinent generator
$G^{(1)}(b)$ which is a tetrahedron of base $b$, that
contains $b(b+1)(b+2)/6$ unit tetrahedrons (see
Fig.~\ref{fig1}).
 The subsequent fractal
stages are constructed self--similarly, by replacing each unit
tetrahedron of the initial generator by a new generator. Thus, to
obtain the $r$th--stage fractal lattice $G^{(r)}(b)$, which we
shall call the $r$th order generator, the recursive process has to
be repeated $(r-1)$ times, so that the complete fractal is
obtained in the limit $r\to\infty$. Fractal dimension $d_f$ of 3d
SG fractal is equal to
\begin{equation}
d_f^{3d}={{\ln [{{(b+2)(b+1)b}/ 6}}]/{\ln b}}\, .
\label{eq:FraktalnaDimenzija3d}
\end{equation}
We assume here that one of the four boundaries of the SG fractal
is impenetrable attractive surface (wall), which is itself a 2d SG
fractal with the fractal dimension
\begin{equation}
d_f^{2d}=\ln[b(b+1)/2]/\ln b\, . \label{eq:FraktalnaDimenzija2d}
\end{equation}

In order to describe both the effect of monomer--monomer
interaction and the effect of attractive (adsorbing) surface, one
should introduce the three Boltzmann factors: $v={\mathrm{
e}}^{-\varepsilon_v/k_BT}$,
$w={\mathrm{e}}^{-\varepsilon_w/k_BT}$, and
$t={\mathrm{e}}^{-\varepsilon_t/k_BT}$, where $\varepsilon_v$ is
the energy corresponding to interaction between two nonconsecutive
neighboring monomers, $\varepsilon_w$ is the energy of a monomer
lying on the adsorbing surface, and $\varepsilon_t$ is the energy
of a monomer in the layer adjacent to the surface (see Fig.~2).
\begin{figure}
\hskip5cm
\includegraphics[scale=0.2]{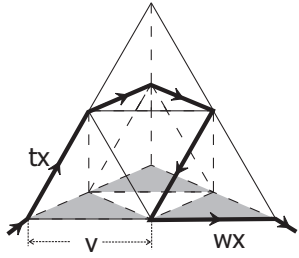}
\caption{ The fractal structure of the $b=2$ 3d SG fractal at the
first stage of construction, with an example of the SAW path. The
shaded  area at the basis of the thetrahedron represents the
adsorption wall. The steps on the adsorbing wall and in the
adjacent layer are weighted by the factors $w=e^{-\epsilon_w/T}$
and $t=e^{-\epsilon_t/T}$, respectively. Here $\epsilon_w$ is the
energy of a monomer lying on the adsorbing wall ($\epsilon_w<0$),
and $\epsilon_t>0$ is the energy  of a monomer that appears in the
layer adjacent to the wall. The Boltzmann factor
$v=e^{-\epsilon_v/T}$ corresponds to the energy of interaction
$\epsilon_v<0 $ between two nonconsecutive neighboring monomers.
The depicted SAW path represent one term in equation
(\ref{eq:A1Gen}) for $r=1$ with $N=5$, $P=4$, $M=1$ and $K=2$.}
 \label{fig2}
\end{figure}
If
we assign the weight $x$ to a single step of the SAW walker, then
the weight of a walk having $N$ steps, with $P$ nearest neighbor
contacts, $M$ steps on the surface, and $K$ steps in the layer
adjacent to the surface, is $x^Nv^Pw^Mt^K$. An arbitrary SAW
configuration can be described, following \cite{Bouchaud}, by
using the five restricted generating functions (see Fig.~3).
  For
$G^{(r)}(b)$, the generating functions, in terms of the
interaction parameters, have the form
\begin{eqnarray}
 A^{(r)}(x,v)=\sum_{N,P}{\mathcal A}^{(r)}(N,P)x^Nv^P,
 \label{eq:AGen}\\
B^{(r)}(x,v)=\sum_{N,P}{\mathcal B}^{(r)}(N,P)x^Nv^P
,\label{eq:BGen}\\
A_1^{(r)}(x,v,w,t)=\sum_{N,P,M,K}{\mathcal
A}_1^{(r)}(N,P,M,K)x^Nv^Pw^Mt^K  ,\label{eq:A1Gen} \\
A_2^{(r)}(x,v,w,t)=\sum_{N,P,M,K}{\mathcal
A}_2^{(r)}(N,P,M,K)x^Nv^Pw^Mt^K  ,\label{eq:A2Gen} \\
B_1^{(r)}(x,v,w,t)=\sum_{N,P,M,K}{\mathcal
B}_1^{(r)}(N,P,M,K)x^Nv^Pw^Mt^K ,\label{eq:B1Gen}
\end{eqnarray}
where the coefficients have the following meanings:
\begin{itemize}
\item
${\mathcal A}^{(r)}(N,P)$ is the number of $N$-step SAWs, lying
completely in the bulk, with $P$ nearest neighbor contacts, and
entering $G^{(r)}(b)$ through one corner vertex, and leaving it
via a second corner vertex,
\item
${\mathcal B}^{(r)}(N,P)$ is the number of $N$-step SAWs,
traversing the $G^{(r)}(b)$ twice, in the bulk, with $P$ nearest
neighbor contacts,
\item
${\mathcal A}_1^{(r)}(N,P,M,K)$ (${\mathcal A}_2^{(r)}(N,P,M,K)$)
is the number of $N$-step SAWs entering the $G^{(r)}(b)$ through a
corner vertex lying on the adsorbing surface, and leaving it via a
second corner vertex on the surface (in the bulk, in the case of
${\mathcal A}_2^{(r)}(N,P,M,K)$), with $P$ nearest neighbor
contacts, $M$ steps in the surface, and $K$ steps in the layer
adjacent to the surface (see Fig.~2), and, finally,
\item
${\mathcal B}_1^{(r)}(N,P,M,K)$ is the number of $N$-step SAWs
going twice through the $G^{(r)}(b)$, with $P$ nearest neighbor
contacts, $M$ steps in the surface, and $K$ steps in the layer
adjacent to the surface.
\end{itemize}
These generating functions (depicted in Fig.~\ref{fig3})
\begin{figure}
\hskip5cm
\includegraphics[scale=0.3]{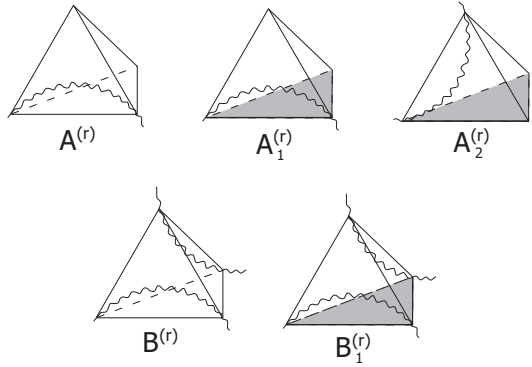}
\caption{ \label{fig3} Schematic representation of the five
restricted generating functions used in describing all possible
polymer configuration within the $r$th stage 3d SG fractal
structure. Thus, for example, $A_1^{(r)}$ represents the SAW paths
that start at one tetrahedron vertex that lies on the adsorption
wall, and exit at the other vertex that also lies on the
adsorption wall. The interior details of the $r$-th order fractal
structure are not shown (they are manifested by the wiggles of the
SAW paths).}
\end{figure}
are parameters in the renormalization group (RG) approach,
and for any $b\geq 2$, RG equations have the form
\begin{eqnarray}
\fl A^{(r+1)}=\sum_{N_A,N_B}a(N_A,N_B) A^{N_A}B^{N_B}  ,
\label{eq:RGA}
\\
\fl B^{(r+1)}=\sum_{N_A,N_B}b(N_A,N_B) A^{N_A}B^{N_B}  ,
\label{eq:RGB}\\
\fl A_1^{(r+1)}=\sum_{N_A,N_B,N_{A_1},N_{A_2},N_{B_1}}
 a_1(N_A,N_B,N_{A_1},N_{A_2},N_{B_1}) A^{N_A}B^{N_B}
A_1^{N_{A_1}}A_2^{N_{A_2}}B_1^{N_{B_1}}  ,  \label{eq:RGA1} \\
\fl A_2^{(r+1)}=\sum_{N_A,N_B,N_{A_1},N_{A_2},N_{B_1}}
 a_2(N_A,N_B,N_{A_1},N_{A_2},N_{B_1}) A^{N_A}B^{N_B}
A_1^{N_{A_1}}A_2^{N_{A_2}}B_1^{N_{B_1}}  , \label{eq:RGA2} \\
\fl B_1^{(r+1)}=\sum_{N_A,N_B,N_{A_1},N_{A_2},N_{B_1}}
 b_1(N_A,N_B,N_{A_1},N_{A_2},N_{B_1}) A^{N_A}B^{N_B}
A_1^{N_{A_1}}A_2^{N_{A_2}}B_1^{N_{B_1}}  , \label{eq:RGB1}
\end{eqnarray}
where we have omitted the superscript $(r)$  on the right-hand
side of the above relations. The self--similarity of the fractals
under study implies that  numbers $a(N_A,N_B)$, $b(N_A,N_B)$,
$a_1(N_A,N_B,N_{A_1},N_{A_2},N_{B_1})$,
$a_2(N_A,N_B,N_{A_1},N_{A_2},N_{B_1})$, and
$b_1(N_A,N_B,N_{A_1},N_{A_2},N_{B_1})$, of the corresponding SAW
configurations within the $G^{(r+1)}(b)$ structure do not depend
on $r$. Starting with the initial conditions
\begin{eqnarray}
A^{(0)}(x,v)=x+2x^2v+2x^3v^3   , \nonumber\\
 B^{(0)}(x,v)=x^2v^4  ,\nonumber\\
A_1^{(0)}(x,v,w,t)=wx+(w^2+t^2)x^2v+2wt^2x^3v^3 ,\nonumber \\
A_2^{(0)}(x,v,w,t)=tx+2twx^2v+2tw^2x^3v^3   , \nonumber\\
B_1^{(0)}(x,v,w,t)=wtx^2v^4 ,\label{eq:PocetniUslovi}
\end{eqnarray}
which correspond to the elementary tetrahedron $G^{(0)}(b)$, one
can iterate RG relations (\ref{eq:RGA})-(\ref{eq:RGB1}) for
various values of interactions $v$, $w$ and $t$, and explore the
phase diagram. This approach (which implies that interactions are
restricted to sites within the elementary tetrahedron $G^{(0)}$
and, moreover, that SAW exits $G^{(r)}$ whenever it reaches its
corner vertex \cite{DharVannimenus}) was applied in
\cite{Bouchaud} for 3d $b=2$ SG fractal. Here we present an
analogous type of analysis for the larger $b$ cases.

The average number of monomers in contact with the adsorption
wall, for SAW spanning a $G^{(r)}(b)$ can be expressed in terms of
the partial derivatives of the generating functions $A_1^{(r)}$
and $A_2^{(r)}$:
\begin{eqnarray}
\fl \langle M^{(r)}\rangle={{\sum_{N,P,M,K}M\left( {\mathcal
A}_1^{(r)}(N,P,M,K)+{\mathcal
A}_2^{(r)}(N,P,M,K)\right)x^Nv^Pw^Mt^K}\over{A_1^{(r)}+A_2^{(r)}}}
\nonumber\\
={w\over{A_1^{(r)}+A_2^{(r)}}} \left( {{\partial
A_1^{(r)}}\over{\partial w}}+{{\partial A_2^{(r)}}\over{\partial
w}}\right)={w\over{A_1^{(r)}+A_2^{(r)}}}
\left(A_{1,w}^{(r)}+A_{2,w}^{(r)}\right)\, , \label{eq:SrednjeM}
\end{eqnarray}
whereas the total average number of monomers can be expressed in
the form
\begin{eqnarray}
\fl \langle N^{(r)}\rangle={{\sum_{N,P,M,K}N\left( {\mathcal
A}_1^{(r)}(N,P,M,K)+{\mathcal
A}_2^{(r)}(N,P,M,K)\right)x^Nv^Pw^Mt^K}\over{A_1^{(r)}+A_2^{(r)}}}
\nonumber\\ ={x\over{A_1^{(r)}+A_2^{(r)}}} \left( {{\partial
A_1^{(r)}}\over{\partial x}}+{{\partial A_2^{(r)}}\over{\partial
x}}\right)={x\over{A_1^{(r)}+A_2^{(r)}}}
\left(A_{1,x}^{(r)}+A_{2,x}^{(r)}\right)\, . \label{eq:SrednjeN}
\end{eqnarray}
From the RG equations (\ref{eq:RGA})-(\ref{eq:RGB1}) one can
obtain recursion relations for the derivatives of the generating
functions in the following matrix form
\begin{equation}
\left( \matrix{A_{1,w}'\cr A_{2,w}'\cr B_{1,w}'}\right)= \mathbf
R_S\left( \matrix{A_{1,w}\cr A_{2,w}\cr B_{1,w}}\right)\, , \quad
\left( \matrix{A_x'\cr B_x'\cr A_{1,x}'\cr A_{2,x}'\cr
B_{1,x}'}\right)= \mathbf R\left( \matrix{A_x\cr B_x\cr A_{1,x}\cr
A_{2,x}\cr B_{1,x}}\right)\, , \label{eq:IzvodiPoxIw}
\end{equation}
where matrices $ \mathbf R_S$ and  $\mathbf R$ are
comprised of partial derivatives of generating functions
$A'$, $B'$, $A_1'$, $A_2'$ and $B_1'$, corresponding to
SAWs spanning the generator $G^{(r+1)}(b)$, in respect to
generating functions $A, B, A_1, A_2$ and $B_1$,
corresponding to $G^{(r)}(b)$:
\begin{displaymath}
 \mathbf R_S=\left( \matrix{
 {{\partial A_1'}\over{\partial A_1}}&{{\partial A_1'}\over{\partial
 A_2}}&{{\partial A_1'}\over{\partial B_1}}\cr
{{\partial A_2'}\over{\partial A_1}}&{{\partial
A_2'}\over{\partial
 A_2}}&{{\partial A_2'}\over{\partial B_1}}\cr
 {{\partial B_1'}\over{\partial A_1}}&{{\partial B_1'}\over{\partial
 A_2}}&{{\partial B_1'}\over{\partial B_1}}}
 \right)\, , \quad
  \mathbf R=\left( \matrix{ {{\partial A'}\over{\partial
A}}&{{\partial A'}\over{\partial
 B}}&0&0&0\cr {{\partial B'}\over{\partial A}}&{{\partial
B'}\over{\partial
 B}}&0&0&0\cr {{\partial A_1'}\over{\partial A}}&{{\partial
 A_1'}\over{\partial B}}&{{\partial A_1'}\over{\partial
 A_1}}&{{\partial A_1'}\over{\partial A_2}}&{{\partial
 A_1'}\over{\partial B_1}}\cr {{\partial A_2'}\over{\partial
 A}}&{{\partial A_2'}\over{\partial B}}&{{\partial
 A_2'}\over{\partial A_1}}&{{\partial A_2'}\over{\partial
 A_2}}&{{\partial A_2'}\over{\partial B_1}}\cr {{\partial
 B_1'}\over{\partial A}}&{{\partial B_1'}\over{\partial
 B}}&{{\partial B_1'}\over{\partial A_1}}&{{\partial
 B_1'}\over{\partial A_2}}&{{\partial B_1'}\over{\partial B_1}}}
 \right)\, . \label{eq:matrice}
 \end{displaymath}
Starting with the initial conditions for the derivatives
\begin{eqnarray}
{{\partial A^{(0)}}\over{\partial x}}=1+2xv+6x^2v^3\,   , \quad
{{\partial B^{(0)}}\over{\partial x}}=2xv^4  ,\nonumber\\
{{\partial A_1^{(0)}}\over{\partial
x}}=w+2(w^2+t^2)xv+6wt^2x^2v^3\,   , \quad {{\partial
A_2^{(0)}}\over{\partial x}}=t+4twxv+6tw^2x^2v^3\,   , \nonumber\\
{{\partial B_1^{(0)}}\over{\partial x}}=2wtxv^4\, , \quad
{{\partial A_1^{(0)}}\over{\partial w}}=x+2wx^2v+2t^2x^3v^3\, ,
\nonumber\\ {{\partial A_2^{(0)}}\over{\partial
w}}=2tvx^2+4twx^3v^3\, , \quad {{\partial B_1^{(0)}}\over{\partial
w}}=tx^2v^4 \label{eq:PocetniUsloviZaIzvode}
\end{eqnarray}
and iterating recursion equations (\ref{eq:IzvodiPoxIw}), it is
possible to establish the relation between the average number of
adsorbed monomers $\langle M^{(r)}\rangle$ (\ref{eq:SrednjeM}) and
the average length of the polymer chain $\langle N^{(r)}\rangle$
(\ref{eq:SrednjeN}) in the limit $r\to\infty$, for various values
of the interaction parameters $v, w$, and $t$.

\section{Exact approach: phase diagram and critical exponents}
\label{treca}

To solve exactly the adsorption problem of a self--interacting SAW
for arbitrary member (for any $b\ge 2$) of the 3d SG family, it is
necessary to find all coefficients that appear in the set
(\ref{eq:RGA})--(\ref{eq:RGB1}) of the RG equations. The $b=2$ case
has been completely analyzed in \cite{Bouchaud}, while in the $b=3$
and $b=4$ cases only the bulk RG equations (\ref{eq:RGA}) and
(\ref{eq:RGB}) have been studied, in \cite{Knezevic} and
\cite{Jelena} respectively. In this paper we make a complete
analysis of the $b=3$ and $b=4$ cases, that is, including the
adsorption RG equations (\ref{eq:RGA1})--(\ref{eq:RGB1}).

\subsection{The $b=3$ SG fractal}

The RG equations (\ref{eq:RGA}) and (\ref{eq:RGB}) for the bulk
parameters $A$ and $B$, found in \cite{Knezevic} for the case of
the $b=3$ SG fractal, have the form

\begin{eqnarray}
\fl A^{(r+1)}=A^3+6 A^4+16 A^5+34 A^6+76 A^7+112 A^8+112 A^9+ 64
A^{10}+ 8 A^4 B+ \nonumber\\
\fl\quad 36 A^5 B+140 A^6 B+292 A^7 B +424 A^8 B+ 332 A^9 B+12 A^3
B^2+12 A^4
B^2+ \nonumber\\
\fl \quad 118 A^5 B^2+ 380 A^6 B^2+ 806 A^7 B^2+664 A^8 B^2+72 A^4
B^3 +352 A^5 B^3+
 \nonumber\\
\fl \quad 704 A^6 B^3 + 1728 A^7 B^3+344 A^4 B^4+ 1568 A^5 B^4+848
A^6
B^4+\nonumber\\
\fl \quad
264 A^4 B^5+3192 A^5 B^5+  320 A^3 B^6\, , \label{eq:A} \\
\fl B^{(r+1)}=A^6+12 A^7+40 A^8+60 A^9+32 A^{10}+28 A^6 B + 88 A^7
B+224 A^8 B+\nonumber\\
\fl\quad 160 A^9 B+ 40 A^6 B^2+496 A^7 B^2 +596 A^8 B^2 + 176 A^5
B^3 +768 A^6
B^3+ \nonumber\\
\fl\quad 1056 A^7 B^3+ 88 A^3 B^4+ 264 A^5 B^4 + 2534 A^6 B^4+
1152 A^4 B^5+\nonumber\\ \fl\quad 1888 A^5 B^5+5808 A^4 B^6+1936
A^3 B^7+ 4308 A^2 B^8 \, . \label{eq:B}
 \end{eqnarray}
Here we give a summary of the analysis of the above set of
equations. For small values of the monomer--monomer interaction
($v<v_\theta$), the extended SAW phase fixed point
$(A_E,B_E)=(0.34196,\;0.02395)$ is reached through the RG
transformations. Linearization of the RG equations in the vicinity
of this fixed point gives only one relevant eigenvalue
$\lambda_{\nu}^E=5.36201$. The mean-squared end-to-end distance
$\langle R_N^2\rangle$ of the $N$--step polymer chain, in general
case, behaves asymptotically (for $N>>1$) as
\begin{equation}
\langle R_N^2\rangle\sim N^{2\nu}, \label{eq:nu}
\end{equation}
where the critical exponent $\nu$ is given by
\begin{equation}\label{exponent-nu}
 \nu={{\ln b}\over{\ln \lambda_{\nu}}}\;,
\end{equation}
which for $b=3$ gives $\nu_E=0.6542$. Starting with $v=v_\theta$,
the RG equations lead to the fixed point
$(A_{\theta},B_\theta)=(0.20717,\;0.43075)$, for which both
eigenvalues  $\lambda_\nu^\theta=8.72308$ and
$\lambda_\alpha^\theta=2.45012$ are relevant. At this fixed point
critical exponent $\nu_\theta=\ln 3/\ln \lambda_\nu^\theta=0.5072$
is smaller than $\nu_E$, which is the manifestation of the
so-called collapse transition. At the collapse transition, the
free energy per site $f$ behaves as
\begin{equation}
f\sim\left|v-v_\theta\right|^{2-\alpha}\, ,  \label{eq:FreeEnergy}
\end{equation}
where the critical exponent $\alpha$ in general case is given by
\begin{equation}\label{alpha}
 \alpha=2-{{\ln\lambda_\nu^\theta}\over{\ln\lambda_\alpha^\theta}}\;
 ,
\end{equation}
and in this specific case it is negative, $\alpha=-0.4170$.
Depending on the value of the one--step weight (fugacity) $x$, for
strong monomer--monomer interactions ($v>v_\theta$) RG equations
(\ref{eq:A}) and (\ref{eq:B}) bring about the trivial fixed point
$(A,B)^*=(0,0)$, for $x<x^*(v)$, or  $(A,B)^*=(\infty,\infty)$ for
$x>x^*(v)$, whereas for $x$ precisely equal to $x^*(v)$, the fixed
point $(A_G,B_G)=(0,\infty)$ is reached. Analyzing the RG
equations in the vicinity of $(A_G,B_G)$, by keeping only the
dominant terms in the right--hand side of (\ref{eq:A}) and
(\ref{eq:B}), one can find $\nu_G=0.48195$. Since the fractal
dimension $d_f^{\mathrm{poly}}=1/\nu_G=2.07491$ of the SAW in this
case is larger than the fractal dimension of the extended SAW (for
$v\leq v_\theta$), but  still less than the fractal dimension of
the underlying lattice  $d_f=2.09590$ one can conclude that fixed
point $(0,\infty)$ corresponds to the 'semi-compact' phase
\cite{Knezevic}. This finding is in  contrast with the results
obtained for polymers on homogeneous lattices and on the $b=2$ 3d
SG fractal lattice \cite{DharVannimenus}, where one finds
$d_f^{\mathrm{poly}}=d_f$.

In order to establish the specific form of the complete set of the
exact RG equations (\ref{eq:RGA})--(\ref{eq:RGB1}), required for
the study of the adsorption problem on the $b=3$ SG fractal, we
have enumerated the requisite SAW configurations, achieving
thereby the pertinent RG coefficients. This procedure is rather
complex, as well as the corresponding set of coefficients, and for
this reason we give them in the Appendix~A. The physical picture
that follows from these RG equations, for various values of the
interaction parameters ($v,t$ and $w$), is unusually rich and we
present it in what follows.

Numerical study of the adsorption RG  equations (given in the
Appendix~A) shows that for $w>1$ (attractive surface) an unbinding
transition appears at a finite temperature only if $t<1$
(repulsive interaction in the layer adjacent to the surface). The
nature of this transition depends on the value of the
monomer--monomer interaction $v$. This situation is analogous to
the previously studied cases of the two--dimensional and
three-dimensional $b=2$ Sierpinski gasket fractals
\cite{Bouchaud}, but the new physical features do appear.

\subsubsection{Extended SAW phase}

For the chosen initial conditions (\ref{eq:PocetniUslovi}) we
found the critical value of monomer--monomer interactions
$v_\theta=2.446161$ (which is different from the value found in
\cite{Knezevic}, because of the slightly different initial
conditions, but which does not affect the overall critical
behavior). For weak monomer--monomer interactions $v<v_\theta$,
and the corresponding critical fugacity $x=x^*(v)$, RG equations
(\ref{eq:A}) and (\ref{eq:B}) for the bulk parameters $A$ and $B$
iterate towards the fixed point $(A_{E},B_{E})$.

Behavior of surface RG parameters ($A_1^{(r)}, A_2^{(r)},
B_1^{(r)}$) depends primarily on the corresponding interaction
parameter $w$. Thus, for weak interactions $w<w_c(t,v)$, and
$x=x^*(v)$, the parameters ($A_1^{(r)}, A_2^{(r)}, B_1^{(r)}$)
tend to $(0,0,0)$, when $r\to\infty$, which indicates that
polymer, being in the extended coil phase, stays away from the
attractive surface. We shall refer to this state as the desorbed
extended (DE) phase, determined by the RG fixed point
\begin{equation}
(A,B,A_1,A_2,B_1)^*=(A_{E},B_{E},0,0,0)\> .
\label{eq:DesorbedSAWFP}
\end{equation}
This behavior changes abruptly at $w=w_c(t,v)$, where $A_1^{(r)},
A_2^{(r)}\to A_{E}$ and $B_1^{(r)}\to B_{E}$. The new behavior is
described by the new fixed point
\begin{equation}
(A,B,A_1,A_2,B_1)^*=(A_{E},B_{E},A_{E},A_{E},B_{E})\>,
\label{eq:SymmetricSpecialFP}
\end{equation}
which is the symmetric special fixed point, that corresponds to
the unbinding transition of the SAW. If $w$ is increased beyond
$w_c(t,v)$, and for the fugacity equal to $x=x^*(v)$, bulk
parameters still approach $(A_{E},B_{E})$, but the surface RG
parameters diverge, implying that critical fugacity $x_c$  is
smaller than $x^*(v)$. A thorough analysis shows that $x_c$
depends on the values of $v,w$, and $t$, whereas RG parameters
flow towards the new fixed point
\begin{equation}
(A,B,A_1,A_2,B_1)^*=(0,0,A_{E}^{2d},0,0)\> ,
\label{eq:AdsorbedSAWFP}
\end{equation}
where $A_{E}^{2d}=0.551147$ is the fixed point for the $b=3$
two--dimensional SG fractal (see \cite{EKM}), meaning that for
strongly attractive surface polymer remains adsorbed (this is the
adsorbed SAW phase).

At the symmetric special fixed point (\ref{eq:SymmetricSpecialFP})
linearized RG equations have two relevant eigenvalues: the larger
$\lambda_\nu^E=5.36201$, already found for the bulk RG equations,
and the smaller $\lambda_S^E=3.32923$. The matrix of the
linearized RG equations is equal to the matrix $\mathbf R$ in
(\ref{eq:IzvodiPoxIw}), which means that
\begin{equation}
\langle N^{(r)}\rangle\sim \left(\lambda_\nu^E\right)^r\, , \qquad
\mathrm{for} \quad r\gg 1\, ,
\end{equation}
whereas the largest eigenvalue of the matrix $\mathbf R_S$ in
(\ref{eq:IzvodiPoxIw}) is equal to $\lambda_S^E$, meaning that
\begin{equation}
\langle M^{(r)}\rangle\sim \left(\lambda_S^E\right)^r\, , \qquad
\mathrm{for} \quad r\gg 1\, .
\end{equation}
Consequently, for $r\to\infty$, it follows
\begin{equation}
\langle M^{(r)}\rangle\sim\langle N^{(r)}\rangle^\phi\> ,
\end{equation}
where the crossover exponent $\phi$ is equal to
\begin{equation}
\phi_E={{\ln\lambda_S^E}\over{\ln\lambda_\nu^E}}=0.7162\, .
\end{equation}
This completes our analysis of the extended polymer phase.

\subsubsection{$\theta$--line ($v=v_\theta$)}

Here we organize our discussion of various phases that appear
along the $\theta$--line in the phase space ($v,w$), for
$v=v_\theta=2.446161$ and for various values of $w$.

For $x=x^*(v_\theta)=0.109683$ and $w<w_\theta(t)$, RG iterations
lead to the fixed point
\begin{equation}
(A,B,A_1,A_2,B_1)^*=(A_\theta,B_\theta,0,0,0)\, ,
\label{eq:DesorbedThetaFP}
\end{equation}
which means that for small values of $w$ polymer remains desorbed,
in the solution, in the form of the $\theta$--chain. On the other
hand, for $w>w_\theta(t)$ critical fugacity $x_c(v_\theta, w,t)$
is less then $x^*(v_\theta)$, and the relevant fixed point is
again  $(0,0,A_{E}^{2d},0,0)$ which describes an adsorbed
two--dimensional SAW (as in the case $v<v_\theta$).

At the critical value $w=w_\theta(t)$, and $x$ still equal to
$x^*(v_\theta)$, RG parameters flow towards the new fixed point
\begin{equation}
(A,B,A_1,A_2,B_1)^*=(A_\theta,B_\theta,A_{1\theta},A_{2\theta},B_{1\theta})\,
, \label{eq:AdsorbedThetaFP}
\end{equation}
whose coordinates $A_{1\theta},A_{2\theta}$, and $B_{1\theta}$
depends on the value of $t$, and accordingly there are five
possible situations that should be analyzed.

First, for $0\leq t< t_1^*=0.1553901$, the new fixed point is
\begin{equation}
(A,B,A_1,A_2,B_1)^*=(A_\theta,B_\theta,A_{E}^{2d},0,0)\, ,
 \end{equation}
that controls the coexistence between the $\theta$--chain in the
bulk and the adsorbed two--dimensional SAW.

Second, we find the fixed point
\begin{equation}
(A,B,A_1,A_2,B_1)^*=(A_\theta,B_\theta,0.5234,0.1637,0.0305)\, ,
 \end{equation}
for the critical value of the interaction $t=t_1^*$, with four
eigenvalues larger than one --- the two bulk values
$\lambda_\nu^\theta$ and $\lambda_\alpha^\theta$, and the two
surface eigenvalues $\lambda_{S1}^\theta=3.96291$, and
$\lambda_{S2}^\theta=1.28383$, which brings about
\begin{equation}
\phi_\theta={{\ln\lambda_{S1}^\theta}\over{\ln\lambda_\nu^\theta}}
=0.6357\, .
\end{equation}

Third, in the interval between the two critical values of the
interaction parameter $t$,
 $t_1^*< t< t_2^*=0.6573781$, the new fixed point is
\begin{equation}
(A,B,A_1,A_2,B_1)^*=(A_\theta,B_\theta,0.19265,0.44713,0.132446)\,
,
 \end{equation}
for which, in addition to the relevant bulk eigenvalues, there is
only one surface relevant eigenvalue $\lambda_S^\theta=3.663797$,
and the crossover exponent is equal to
\begin{equation}
\phi_\theta={{\ln\lambda_S^\theta}\over{\ln\lambda_\nu^\theta}}
=0.5995 \, .
\end{equation}

Fourth, for the second critical $t_2^*$, the RG parameters flow
towards the symmetric fixed point
\begin{equation}
(A,B,A_1,A_2,B_1)^*=(A_\theta,B_\theta,A_{\theta},A_\theta,B_\theta)\,
,
\end{equation}
where, as in the case of the first critical value $t_1^*$, one
finds two relevant surface eigenvalues
$\lambda_{S1}^\theta=5.368208$ and $\lambda_{S2}^\theta=1.718244$,
wherefrom one obtains the crossover exponent
\begin{equation}
\phi_\theta={{\ln\lambda_{S1}^\theta}\over{\ln\lambda_\nu^\theta}}
=0.7758\, .
\end{equation}

Fifth, for $t_2^*< t< 1$ the following fixed point
\begin{equation}
(A,B,A_1,A_2,B_1)^*=(A_\theta,B_\theta,0,0,\infty)
 \end{equation}
is reached. By keeping the dominant terms on the right--hand side
of the RG equations (\ref{eq:A1}--\ref{eq:B1}), for the surface
parameters $A_1, A_2$, and $B_1$ (see Appendix A), the RG
equations attain the approximate form
\begin{eqnarray}
\fl A_1' \approx
(128A_\theta^2+124A_\theta^3+292A_\theta^4+264A_\theta^2B_\theta+
944A_\theta^3B_\theta+320A_\theta B_\theta^2)A_1^2B_1^4
=c_1A_1^2B_1^4\, ,\nonumber\\
\fl B_1'
\approx(472A_\theta^4+1452A_\theta^3B_\theta+1452A_\theta^2B_\theta^2+4308A_\theta
B_\theta^3)A_1B_1^5=c_2A_1B_1^5\, , \nonumber\\
\fl A_2'\approx c_3A_2\, ,
\end{eqnarray} where $c_1\approx 6.26$,
$c_2\approx 89.32$, and $c_3\approx 0.127$. Introducing the new
variable $y_1=B_1A_1^{1/4}$ we obtain the tractable form of the
approximate RG equations
\begin{equation}
A_1'=c_1y_1^4A_1\, , \quad y_1'=c_1^{1/4}c_2y_1^6\, , \quad
A_2'=c_3A_2\, ,
\end{equation}
which have the fixed point $(0,0,y_1^*=c_1^{-1/20}c_2^{-1/5})$.
Linearizing these RG equations in the vicinity of the fixed point,
one relevant eigenvalue $\lambda_S^\theta=6$ is found, with the
corresponding crossover exponent $\phi_\theta$
\begin{equation}
\phi_\theta={{\ln\lambda_S^\theta}\over{\ln\lambda_\nu^\theta}}
=0.8272\,.
\end{equation}

\subsubsection{Semi-compact regime ($v>v_\theta)$}

For values of the monomer--monomer interaction parameter $v$
larger than $v_\theta$, and for the fugacity $x$ equal to the bulk
critical value $x^*(v)$, the bulk RG parameters ($A, B$) flow
towards $(0,\infty)$, while critical behavior of the surface RG
parameters ($A_1, A_2, B_1$) depends on the values of both surface
interaction parameters $w$ and $t$. In particular, for any value
of $t$ between 0 and 1, there is a critical value $w_c(t,v)$ such
that ($A_1, A_2, B_1$) flows towards $(0,0,0)$, for $w<w_c(t,v)$,
whereas for $w$ precisely equal to $w_c(t,v)$ one observes
\begin{equation}\label{semicompact1}
(A,B,A_1,A_2,B_1)\to \left\{ \matrix{(0,\infty,A^{2d}_{SAW},0,0)\,
, & v_\theta<v<v_\theta+\epsilon \cr (0,\infty,0,0,\infty)\, ,&
v>v_\theta+\epsilon \cr} \right.
\end{equation}
for $t<t^*\approx 0.156$, and
\begin{equation}
(A,B,A_1,A_2,B_1)\to (0,\infty,0,0,\infty)\, ,
\end{equation}
 for $t>t^*$ and all $v>v_\theta$.

The fixed points $(0,\infty,0,0,0)$,
$(0,\infty,A^{2d}_{SAW},0,0)$, and $(0,\infty,0,0,\infty)$,
correspond to the desorbed semi--compact polymer chain (globule),
to the coexistence between the globule in the bulk and the
adsorbed 2d polymer, and to the adsorbed globule, respectively.
Finally, for values of the parameter $w$ larger than the critical
value $w_c(t,v)$, the critical fugacity $x_c=x^*(v,w,t)$ is
smaller than its bulk critical value $x^*(v)$, and RG parameters
iterate towards  the fixed point $(0,0,A^{2d}_{SAW},0,0)$, which
corresponds to the fully adsorbed polymer.

The adsorbed globule phase can be analyzed by keeping the dominant
terms on the right-hand side of the RG equations (\ref{eq:A}),
(\ref{eq:B}), and (\ref{eq:A1})--(\ref{eq:B1}), in the vicinity of
the fixed point $(0,\infty,0,0,\infty)$. Accordingly, one can
obtain the following approximate equations:
\begin{eqnarray}
A'&\approx& 320 A^3 B^6\, , \quad
 B'\approx 4308 A^2 B^8\, ,\quad
A_1'\approx 320 A A_1^2 B^2 B_1^4\, ,\nonumber \\ A_2'&\approx& 44
A A_1^3 B B_1^3\, ,\quad B_1'\approx 4308 A A_1 B^3 B_1^5\, .
\label{eq:aproxxb3}
\end{eqnarray}
Introducing new variables
\begin{equation}
y=AB^z\, , \quad  y_1=A_1/A   \, ,\quad y_2=A^q/A_2\, , \quad
y_3=B_1 A^z\, ,
\end{equation}
where
\begin{equation}
z={{\sqrt{73}-5}\over{12}}   \, ,\quad
q={{19+\sqrt{73}}\over{12}}\, ,
\end{equation}
equations (\ref{eq:aproxxb3}) transform into more tractable form
\begin{eqnarray}
A'&=& 320  A^{3-z}y^6\, , \quad y'= 320^z  4308 \, y^{8+6z}\, ,
\label{eq:Ay} \\
 y_1'&=&{{y_1^2 y_3
^4}\over{y^4}}\, , \quad  y_3'=  320^z  4308 \, y_1 y^{6z+3}
y_3^5\, ,\label{eq:y1y3}\\
y_2'&=&{{320^q}\over{44}}{{y^{6q-1}}\over{y_1^3 y_3^3}}\, .
\label{eq:y2}
\end{eqnarray}
The new equations (\ref{eq:Ay}) have fixed point $A^*=0$,
$y^*=(4308\times320^z)^{-1/(7+6z)}$, with one relevant bulk
eigenvalue $\lambda_\nu^G=8+6 z=(11+\sqrt{73})/2$ (see
\cite{Knezevic}). Inserting $y^*$ into equations (\ref{eq:y1y3})
one finds that corresponding equations for the fixed point
$(y_1^*,y_3^*)$ are mutually linearly dependent. More precisely,
the only fact that springs from these equations is the relation
\begin{equation}
y^*=(y_1^*)^{1/4} y_3^*\, .
\end{equation}
For various values of $t$, large enough $v$ and corresponding bulk
critical fugacity $x^*(v)$, and the critical value $w_c(t,v)$,
point $(y_1,y_3)$ tends to different fixed points $(y_1^*,y_3^*)$,
but the above relation stays satisfied. Obviously, knowing
$(y_1^*,y_3^*)$, from equation (\ref{eq:y2}) one can calculate the
fixed value $y_2^*$, which is also different for various $t$ and
$v$, but on the other hand is in excellent agreement with the
value obtained via explicit iteration of the expression $A^q/A_2$.

Linearizing equations (\ref{eq:y1y3}), (\ref{eq:y2}) around the
corresponding fixed point $(y_1^*,y_2^*,y_3^*)$ the second
relevant eigenvalue $\lambda_S^G=6$ is found, so that the
crossover exponent $\phi$ is equal to
\begin{equation}
\phi_G=\displaystyle{\frac{\ln 6}{\ln (\frac{11 +
{\sqrt{73}}}{2})}}=0.786\, .
\end{equation}
This result is in agreement with the value estimated via
direct numerical analysis, performed using equations
(\ref{eq:SrednjeM}) and (\ref{eq:SrednjeN}), as explained
in Section~\ref{druga}. One should also observe that
obtained value of $\phi_G$ is slightly larger than
${{d_f^{2d}}/{d_f^{3d}}}=0.7781$, which has been predicted
in \cite{Bouchaud} for SAW in the compact phase. Of course,
due to the topological frustration,  SAW on 3d $b=3$
Sierpinski fractal lattice can not, even for large
monomer---monomer interaction, have a compact configuration
(see \cite{Knezevic}). Instead, it is in the semi-compact
phase, in which its fractal dimension is less than the
fractal dimension of the lattice (although larger than
fractal dimension of SAW for $v\leq v_\theta$).

\subsection{3d $b=4$ SG fractal}

As the scaling parameter $b$ increases the number of possible
polymer configurations on the 3d SG fractal lattices quickly
grows. The RG equations, for $b=4$, for the bulk RG parameters $A$
and $B$ have been found and analyzed by Mari\v ci\' c and
Elezovi\' c--Had\v zi\' c (see \cite{Jelena} and Appendix B). The
conclusion has been reached that qualitatively the physical
picture of the polymer behavior in the bulk, for $b=4$, is similar
to the cases $b=2$ and $b=3$. There are three possible phases in
which polymer can reside, which we review in the following
paragraph.

For small values of the monomer--monomer interaction
$v<v_\theta=2.33187$ polymer is in the swollen state. The
fixed point $(A,B)^*=(A_E,B_E)=(0.2899, 0.0122)$ is reached
for any $v<v_\theta$ and for the corresponding critical
fugacity $x=x^*(v)$. Linearizing RG equations around this
fixed point one gets single relevant eigenvalue
$\lambda_\nu^E=8.6924$, and the concomitant critical
exponent $\nu$ is equal to $\nu_{E}=\ln
b/\ln\lambda_\nu^E=0.6410$. On the other hand, the
low--temperature fixed point $(A,B)^*=(A_G,B_G)=(0,
22^{-1/3})$ is reached when the RG iteration starts with
$v>v_\theta$. In this case, there is one relevant
eigenvalue $\lambda_\nu^G=16$, and the end-to-end critical
exponent is $\nu_G=1/2$. The fractal dimension of the chain
$d_f^{\mathrm{poly}}=1/\nu_G=2$ is less than the fractal
dimension of the lattice $d_f=\ln 20/\ln 4\approx 2.16$.
This means that polymer is in the semi--compact phase for
strong monomer--monomer interactions, as in the $b=3$ case,
and in contrast with the cases of polymer on the $b=2$
fractal and on the Euclidean lattices. Finally, when
$v=v_\theta$, the tricritical fixed point
$(A,B)^*=(A_\theta,B_\theta)=(0.1929, 0.3388)$, that
corresponds to the collapse transition, is reached. In this
case there are two relevant eigenvalues,
$\lambda_\nu^\theta=15.4230$ and
$\lambda_\alpha^\theta=5.5357$. Consequently, the critical
exponent $\nu$ is equal to $\nu_\theta=\ln
b/\ln\lambda_\nu^\theta=0.5067$. In addition, we have
obtained the specific heat critical exponent
$\alpha=0.4012>0$  which reveals the singular behavior,
which was observed in the case $b=2$ but not in the case
$b=3$.

Recursion relations for the surface RG parameters $A_1$, $A_2$,
and $B_1$ are cumbersome (each of the corresponding equations has
more than 3000 terms) and we are not going to quote them here (but
they are available upon the request to the authors). Detailed
numerical analysis of these RG equations shows that, depending on
the value of interaction parameters $v$, $w$, and $t$, the
following phases and the corresponding fixed points are
accessible:
\begin{itemize}
\item
extended desorbed phase, $(A_{E},B_{E},0,0,0)$, for $v<v_\theta$
and  $w<w_c(t,v)$,
\item
semi-compact desorbed phase, $(0,B_G,0,0,0)$, for $v>v_\theta$ and
$w<w_c(t,v)$,
\item
desorbed $\theta$--chain, $(A_\theta,B_\theta,0,0,0)$, for
$v=v_\theta$ and $w<w_\theta(t)$,
\item
fully adsorbed chain, $(0,0,A^{2d}_{SAW},0,0)$, for $w>w_c(t,v)$,
where $A^{2d}_{SAW}=0.5063$ is the fixed point for the 2d $b=4$ SG
fractal (see \cite{EKM}),
\item
coexistence of the globule in the bulk and the 2d adsorbed polymer
chain, $(0,B_G,A^{2d}_{E},0,0)$, for
$v_\theta<v<v_\theta+\epsilon$ (where $\epsilon$ is a small
positive number whose specific value depends on $t$) and
$w=w_c(t,v)$,
\item
 surface attached extended chain, special symmetric fixed point,
$(A_{E},B_{E},A_{E},A_{E},B_{E})$, for $v<v_\theta$ and
$w=w_c(t,v)$, with the relevant eigenvalues $\lambda_\nu^E$
and $\lambda_S^E=4.4533$ that give the crossover exponent
$\phi_E=\ln\lambda_S^E/\ln\lambda_\nu^E=0.6907$,
\item
surface attached globule, $(0,B_G,0,0,B_G)$, for
$v>v_\theta+\epsilon$ and $w=w_c(t,v)$, with the relevant
eigenvalues $\lambda_\nu^G$ and $\lambda_S^G=9$ and
$\phi_G=0.7924$,
\item
surface attached $\theta$--chain, multi-critical point
$(A_\theta,B_\theta,A_{1\theta},A_{2\theta},B_{1\theta})$, for
$v=v_\theta$ and $w=w_\theta(t)$, with
\begin{equation}
\fl (A_{1\theta},A_{2\theta},B_{1\theta})=\left\{ \matrix{
(0.2275, 0.3659, 0.0433),& t<t^*\cr (A_\theta,A_\theta,B_\theta),
& t=t^*=0.9577\cr (0,0, 0.3643), & t^*<t<1} \right.
\end{equation}
for which one finds
\begin{equation}
\fl \lambda_S^\theta=\left\{ \matrix{ 4.9036, & t<t^*\cr 8.4078, &
t=t^*\cr 9 , & t^*<t<1}
 \right.
\quad \mathrm{and} \qquad \phi_\theta=\left\{ \matrix{
0.5811, & t<t^*\cr 0.7782, & t=t^*\cr 0.8031, & t^*<t<1}
 \right.
\end{equation}
\end{itemize}

In the last part of this section we summarize, in
Table~\ref{tabela-suki},
\Table{\label{tabela-suki}  Fixed points
and critical exponents $\nu$ and $\phi$ for different
self--avoiding walk phases on 3d Sierpinski fractals, obtained via
exact renormalization group approach. Values for
 the bulk fixed
point $(A^*,B^*)$ and the critical exponent $\nu$ for $b=2$
and 3 fractals were obtained in \cite{DharVannimenus} and
\cite{Knezevic} respectively, whereas the surface bulk
point $(A_1^*,A_2^*,B_1^*)$ and the crossover exponent
$\phi$ were previously known  for the $b=2$ fractal only
\cite{Bouchaud}.}
 \begin{tabular}{@{}lllllllll}
    \br
    \ms
 \centre{9}{extended chain phase}\\
 \ms
     $b$ & & $A^*$ & $B^*$ & $\nu_E$   & $A_1^*$ & $A_2^*$ & $B_1^*$ & $\phi_E$
    \\\mr
    2 & & 0.4294 & 0.0499 & 0.6740 & 0.4294 & 0.4294 &0.0499  & 0.7481
    \\
    3 & & 0.3491 & 0.0239 & 0.6542& 0.3491 & 0.3491 & 0.0239 & 0.7162
    \\
    4 & & 0.2899 & 0.0122 & 0.6411 & 0.2899 & 0.2899 & 0.0122 & 0.6907 \\
  \br
  \ms
  \centre{9}{$\theta$--chain} \\
  \ms
$b$& & $A^*$ & $B^*$ & $\nu_\theta$ & $A_1^*$ & $A_2^*$ & $B_1^*$
& $\phi_\theta$
\\ \mr
    2& & 1/3 & 1/3 & 0.5923 & 0.4477 & 0.4528 & 0.0815 & 0.6264 \\
 &&&&&\crule{4}\\
    3& & 0.2071 & 0.4307 & 0.5072 & \begin{tabular}{@{}l}
      0.5234 \\
      0.1926 \\
      0.2071 \\
      0 \\
    \end{tabular} & \begin{tabular}{@{}l}
      0.1637 \\
      0.4471 \\
      0.2071 \\
      0 \\
    \end{tabular} & \begin{tabular}{@{}l}
      0.0305 \\
      0.1324 \\
      0.4307 \\
      $\infty$ \\
    \end{tabular} & \begin{tabular}{@{}l}
      0.6357 \\
      0.5995 \\
      0.7758 \\
      0.8272 \\
    \end{tabular} \\
    &&&&&\crule{4}\\
    4& & 0.1929 & 0.3388 & 0.5067 & \begin{tabular}{@{}l}
      0.2275 \\
      0.1929 \\
      0 \\
    \end{tabular} & \begin{tabular}{@{}l}
      0.3659 \\
      0.1929 \\
      0 \\
    \end{tabular} & \begin{tabular}{@{}l}
      0.0433 \\
      0.3388 \\
      0.3643 \\
    \end{tabular} &
\begin{tabular}{@{}l}
      0.5811 \\
      0.7782 \\
      0.8301 \\
    \end{tabular}
     \\
   \br
   \ms
  \centre{9}{globular phase}\\
  \ms
$b$& & $A^*$ & $B^*$ & $\nu_G$ & $A_1^*$ & $A_2^*$ & $B_1^*$ &
$\phi_G$
    \\ \mr
    2 && 0 & $22^{-1/3}$ & 1/2 & 0 & 0 & $22^{-1/3}$ & 0.7925 \\
    3 && 0 & $\infty$ & $0.4819$ & 0 & 0 & $\infty$ & 0.7860 \\
    4 && 0 & $22^{-1/3}$ & 1/2 & 0 & 0 & $22^{-1/3}$ & 0.7925 \\
 \br
\end{tabular}
\endTable
the numerical results obtained via the exact RG approach, for the
3d SG fractals $b=2,3,4$, and provide the relevant discussion of
the pertinent phase diagrams. One should notice that we first give
the fixed point values of the bulk parameters $A^*$ and $B^*$ and
the accompanying end--to--end critical exponent $\nu$. For each
$b$ there are three fixed points, which correspond to the three
bulk phases --- the extended polymer phase, the $\theta$--chain
phase, and the globular (collapsed) phase. Furthermore, for a
given $b$, one may observe that $A^*$ decreases, while $B^*$
increases, with increasing of the monomer-monomer interaction $v$,
and consequently $\nu$ decreases. On the other hand, when $b$
increases $\nu$ decreases for the extended phase and the
$\theta$--chain phase, whereas in the case of the globular phase
$\nu$ does not display a monotonic behavior. Besides, in the
globular phase, for $b=3$ and $b=4$ the exponent $\nu$ is larger
than the reciprocal of the fractal dimension of the underlying
lattices, which implies that the globular phase is not compact.
This is in contrast with the $b=2$ case (as well as, in contrast
with the Euclidean lattices), where $\nu=1/d_f$.

In Table~\ref{tabela-suki} we also give the fixed point values of
the surface RG parameters $A_1^*,A_2^*,B_1^*$ and the
corresponding values of the crossover exponent $\phi$, for the
unbinding transitions from adsorbed polymer phase to various
desorbed phases. The relevant phase diagram is given in Fig.~4,
for the $b=3$ SG fractal (a similar phase diagram for $b=2$ was
obtained in \cite{Bouchaud}, whereas in the course of this work we
obtained a diagram, of the same type, for $b=4$). In Fig.~4, the
unbinding transitions are represented by the curve that separates
the ``adsorbed chain" region from the ``desorbed chain" regions.
On this curve lies the multi-critical point, so that the part of
the curve for smaller values of the parameter $v$ ($v<v_\theta$)
corresponds to the extended attached chain phase, while for larger
values of $v$ ($v>v_\theta$) the curve corresponds to the attached
semi-compact phase. The fixed point that defines the extended
attached chain phase is a symmetric special fixed point, that is,
$A_1^*=A_2^*=A^*$ and $B_1^*=B^*$, with the corresponding
crossover exponent which decreases with increasing of the scaling
parameter $b$ (see the top part of the Table~\ref{tabela-suki}).
Here we would like to mention that this behavior of the crossover
exponent will be demonstrated for larger $b$ as well, using the
MCRG method (see the next section of this paper).

Continuing our discussion guided by the results presented
in Table~\ref{tabela-suki}, we focus now on the attached
globular phase ($v>v_\theta$). For this phase, the fixed
point parameters satisfy the relations $A_1^*=A_2^*=A^*=0$
and $B_1^*=B^*$, where $B^*=22^{-1/3}$ for $b=2$ and $b=4$,
while $B^*=\infty$ for $b=3$. It is interesting to observe
that the critical exponents $\nu$ and $\phi$ are the same
for the fractals with $b=2$ and $b=4$, while for $b=3$
these two exponents are somewhat smaller. Besides, we
should point out that only for $b=2$ the crossover exponent
$\phi$ is equal to the ratio $d_f^{2d}/d_f^{3d}$ which is
related to the fact that only for $b=2$ the globular phase
is compact. Finally, by comparing values of the crossover
exponent $\phi$, we may observe that in the attached
globular phase the number of adsorbed monomers is
relatively larger than in the corresponding Euclidean
phases ($\phi=2/3$).

\begin{figure}
\hskip3cm
\includegraphics[scale=0.5]{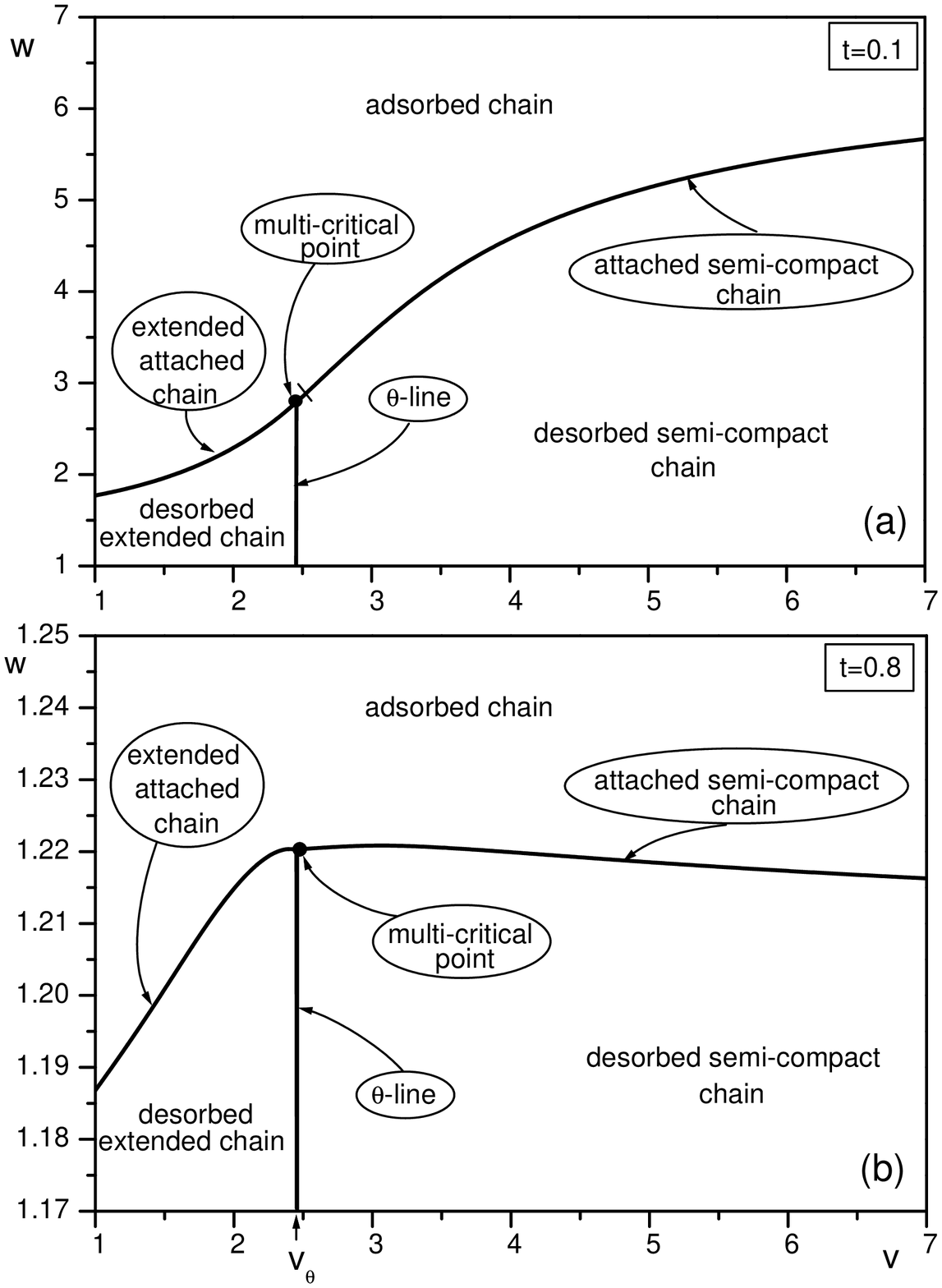}
\caption{Phase diagram in the case of the $b=3$ SG fractal for two
different values of the interaction parameter $t$: (a) $t=0.1$ and
(b) $t=0.8$. For $t=0.1$, close to the multi-critical point on the
critical line exists a small region
($v_\theta<v<v_\theta+\epsilon$) of the first order phase
transitions (where the globule in the bulk and the adsorbed chain
coexist) which ends at the point depicted by the small dash.
Beyond the dash ($v>v_\theta+\epsilon$) the transition at the
critical line is continuous, and corresponds to the attached
semi--compact chain.}
 \label{fig4}
\end{figure}

Results obtained for the multi-critical points for the fractals
with $b=3$ and $b=4$ (see the middle right part of Table I) appear
to be dependent on the parameter $t$ that describes energy of a
monomer in the layer adjacent to the adsorbing boundary. This is
manifested by several possible fixed points, in contrast to the
case $b=2$. Particularly, the position of the multi-critical
point, as well as the shape of the critical line $w_c(v)$ (that
separates the adsorbed phase from the desorbed phases), depend on
the particular value of $t$ (see Fig.~\ref{fig4}).
For the case $b=3$ there are four fixed points that correspond to
the surface attached $\theta$-chain, two of which  (the first and
the third one) appear to be very repulsive (each with four
eigenvalues larger than one) and can be approached for the
critical values $t_1^*=0.1553$ and $t_2^*=0.6573$ of the parameter
$t$. The latter fixed point appears to be symmetric
($A_1^*=A_2^*=A^*=A_\theta$ and $B_1^*=B^*=B_\theta$) and
accordingly is the only point that describes the isotropic chain.
The other two fixed points, the second and the fourth, are less
repulsive (each of them has three eigenvalues larger than one),
and can be reached for any value of the parameter $t$ in the
intervals $t_1^*<t<t_2^*$ and $t_2^*<t<1$, respectively.

Contrary to the expectation, that may be formed on the
findings presented for the cases $b=2$ and $b=3$, that
number of attainable multi-critical points will continue to
increase with increasing $b$, in the $b=4$ case only three
multi-critical points can be reached. Indeed, for $b=4$
there is only one critical value $t^*=0.9577$ of the
parameter $t$ such that for $0<t<t^*$ and for $1>t>t^*$,
two different nonsymmetric fixed points (one for each of
the two intervals) with three eigenvalues larger than one
are reached. On the other hand, highly repulsive symmetric
point, with four eigenvalues larger than one, can be
approached only for $t=t^*$. Each one of the three
multi-critical points describes its pertinent surface
attached $\theta$--chain which is manifested by the facts
that $A_2^*>A_1^*>A^*=A_\theta$ and $B_1^*<B_\theta$ for
$t<t^*$, $A_1^*=A_2^*=0$ and $B_1^*<B_\theta$ for $t>t^*$,
meaning that the corresponding two phases are anisotropic,
whereas for the critical value $t=t^*$ the surface attached
$\theta$-chain is isotropic. Finally, one can notice that
crossover exponent $\phi$ increases with $t$ (which was not
the case for $b=3$), and moreover that the particular
values of $\phi$ for $b=3$ and $b=4$, in their relevant
intervals $1>t\geq t_2^*$ and $1>t\geq t^*$, are rather
close.

\section{Monte Carlo renormalization group calculation}
\label{mcrg}

In this section we are going to apply the Monte Carlo
renormalization group (MCRG) method to calculate the critical
exponents $\nu$ end $\phi$ for 3d SG fractals with $b\ge 5$.
First, we shell present the MCRG calculation of the critical
exponent $\nu$. In order to find the SAW critical exponent $\nu$,
in the bulk phase, we should determine the nontrivial fixed points
of the RG transformations (\ref{eq:RGA}) and (\ref{eq:RGB}), and
then we should solve the corresponding eigenvalue problem of the
linearized RG transformations, that is, we should solve the
equation
\begin{equation}\label{eq:det1}
\left| \matrix{ ({{\partial A'}\over{\partial
A}}-\lambda_\nu)&{{\partial A'}\over{\partial
 B}}\cr
{{\partial B'}\over{\partial A}}&({{\partial B'}\over{\partial
 B}}-\lambda_\nu)\cr}
 \right|^*=0 \, ,
\end{equation}
where the asterisk means that all derivatives should be taken at
the corresponding fixed point and the superscript prime is used
instead of the superscript $(r+1)$. Knowing the relevant
eigenvalue
 $\lambda_\nu$, we can determine the
critical exponent $\nu$ using the formula (\ref{exponent-nu}). To
learn a specific value of $\nu$, for a given $b$, one should first
find the coefficients $a(N_A,N_B)$, and $b(N_A,N_B)$ in the RG
equations (\ref{eq:RGA}) and (\ref{eq:RGB}). As it was detailed in
the previous sections, it has been possible to calculate the exact
values of these coefficients only for $b\le 4$. Thus, to get an
entire sequence of values of $\nu$ for $b\ge 5$, we are going to
circumvent the problem of explicit determination of the exact
coefficients in the RG equations by applying the MCRG technique.

The MCRG method, allows direct calculation of derivatives that
appear in the eigenvalue equation (\ref{eq:det1}). It starts by
locating the bulk nontrivial fixed points, which requires, at the
beginning, implementation of a MC simulation of the SAWs for a
chosen initial set of values $(A_0, B_0)$. In other words, we let
the walker start his walking,  at one fixed corner of the fractal
generator and record the other corner, at which it leaves the
generator, together with recording the total numbers $N_A$ and
$N_B$, of crossings of the $A$, or $B$, type  through the
elementary tetrahedron. The SAW walker crosses an elementary
tetrahedron in the $A$ (or $B$) way (see Fig.~\ref{fig3}) with the
weight (probability) $A_0$, and $B_0$, respectively. We repeat
this MC simulation $L$ times, for the same set $(A_0, B_0)$. Thus
we find how many times the walker has passed through the generator
in the $A$ (or $B$) way and by dividing the corresponding numbers
by $L$ we get the values of the functions (\ref{eq:RGA}) and
(\ref{eq:RGB}), denoted here by $A'(A_0,B_0)$ and $B'(A_0,B_0)$.

In this way we get the value  of the sums (\ref{eq:RGA}) and
(\ref{eq:RGB}) without specifying the coefficients  $a(N_A,N_B)$,
and $b(N_A,N_B)$. Then, the subsequent values $A_n$ and $B_n$
$(n\ge 1)$, at which the MC simulation should be performed, can be
found by using the generalized ``homing" procedure
\cite{Redner,zivic1,zivic5}, which can be terminated at the stage
when the differences $A_n -A_{n-1}$ and $B_n -B_{n-1}$ become less
than the statistical uncertainties associated with $A_{n-1}$ and
$B_{n-1}$, respectively. Consequently, fixed point $(A^*,B^*)$ can
be identified with the last value $(A_n,B_n)$ found in this
procedure.

Having learnt the fixed point, we need to solve the eigenvalue
equation (\ref{eq:det1}) in order to find the critical exponent
$\nu$ via the formula (\ref{exponent-nu}). Thus, we should find
the partial derivatives ${\partial{Y'}/\partial{X}}$ (where
$X,Y\in \{A,B\}$) at the fixed point. For instance, starting with
(\ref{eq:RGA}) (in the notation in which the superscript prime is
used instead of the superscript $(r+1)$) and by differentiating it
with respect to $A$ we get
\begin{equation}\label{eq:1}
{\partial{A'}\over\partial{A}}= \sum_{N_A,N_B} N_A\, a(N_A,N_B)
{\left(A\right)}^{N_A-1} {\left(B\right)}^{N_B}\>.
\end{equation}
 Treating now $A'$ as the grand
canonical partition function, for the ensemble of all possible SAWs
that start at one fixed corner of the fractal generator and leave
it at the other one, we can write the corresponding ensemble
average
\begin{equation}\label{eq:2}
\langle N_A(A,B)\rangle_{A'}= {1\over A'} \sum_{N_A,N_B} N_A\,
a(N_A,N_B) {\left(A\right)}^{N_A} {\left(B\right)}^{N_B}\>,
\end{equation}
which can be directly measured in a MC simulation. Finally,
comparing the last two equations, we can express one of the
requisite derivatives in terms of the measurable quantity
\begin{equation}\label{eq:3}
{\partial{A'}\over\partial{A}}={A'\over A} \langle
N_A(A,B)\rangle_{A'}\>.
\end{equation}

 In a similar way one can find
additional three derivatives, so that we can write the general
formula
\begin{equation}\label{eq:4}
{\partial{Y'}\over\partial{X}}={Y'\over X} \langle
N_X\rangle_{Y'}\>,
\end{equation}
 where $X$ and $Y$ stand for any pair
of quantities from the set $\{ A, B\}$.
 In this way we can learn, through the MC
simulations, the partial derivatives that appear in the eigenvalue
equation (\ref{eq:det1}).
Consequently, calculating the above derivatives at the
 fixed point and solving the eigenvalue equation
(\ref{eq:det1}) we obtain
\begin{equation}\label{lambdani}
 \lambda_\nu={\langle N_A\rangle_{A'}^* +\langle
N_B\rangle_{B'}^* \over 2}+ \sqrt{\left({\langle N_A\rangle_{A'}^*
-\langle N_B\rangle_{B'}^* \over 2}\right)^2+ {\langle
N_A\rangle_{B'}^* \langle N_B\rangle_{A'}^* }}\>,
\end{equation}
which means that $\lambda_\nu$ has been expressed in terms of
quantities that all are measurable through the Monte Carlo
simulations. Accordingly, we can learn  the value of the critical
exponent $\nu$ through the formula (\ref{exponent-nu}). We have
applied this technique for a sequence of of the 3d SG fractals and
in Table~\ref{tabela1} we present our findings for $2\leq
b\leq40$, for the extended chain (in the bulk phase) fixed point
$(A_{E},B_{E})$  together with the related critical exponent
$\nu_E$. As one can see from Table~\ref{tabela1} the fixed point
values for $B_E$, decrease much faster then the values for $A_E$,
when $b$ increases. This means that we may neglect parameter
$B^{(r)}$ (compared with $A^{(r)}$) for larger $b$, that is for
$b>12$. In order to estimate the influence of the parameter
$B^{(r)}$ on the values of the critical exponent $\nu$, we
calculated, by the MCRG method, critical exponent $\nu$ for $b=12$
with $B_E=0$. We obtained the following result
$\nu=0.5989\pm0.0003$, which deviate 0.03\% from the value
$\nu=0.5987\pm0.0003$ (see Table~\ref{tabela1}) found using the
nonzero value of $B^{(r)}$. Concerning the analogous results for
the collapse transitions ($\theta$ critical point) in the bulk
phase, we have to point out that the corresponding point
$(A_\theta,B_\theta)$ cannot be located because the initial part
of the applied technique (the ``homing procedure") does not
possess the needed convergence.

We now apply the MCRG method to find the crossover critical
exponent $\phi$ (which determines the number of adsorbed monomers)
for the polymer phase, on the 3d SG fractals, described by the
symmetric fixed point (\ref{eq:SymmetricSpecialFP}). To this end,
we have to solve the eigenvalue problem for the second part (that
starts with the third equation) of the RG transformations
(\ref{eq:RGA})--(\ref{eq:RGB1}), which reduces to solving of the
equation
\begin{equation}\label{eq:det2}
{\left|\matrix{ \left({\strut \displaystyle \partial {A'_1}\over
\strut\displaystyle \partial A_1} -\lambda_\phi\right) &
{\strut\displaystyle \partial {A'_1}\over \strut\displaystyle
\partial A_2}&
{\strut\displaystyle \partial {A'_1}\over \strut\displaystyle
\partial B_1} \cr
{\strut\displaystyle \partial {A'_2}\over \strut\displaystyle
\partial A_1}&
\left({\strut \displaystyle \partial {A'_2}\over
\strut\displaystyle
\partial A_2} -\lambda_\phi\right) &
{\strut\displaystyle \partial {A'_2}\over \strut\displaystyle
\partial B_1} \cr
{\strut\displaystyle \partial {B'_1}\over \strut\displaystyle
\partial A_1}&
{\strut\displaystyle \partial {B'_1}\over \strut\displaystyle
\partial A_2}&
\left({\strut \displaystyle \partial {B'_1}\over
\strut\displaystyle
\partial B_1} -\lambda_\phi\right) \cr}
\right|}^*=0\>.
\end{equation}
Here the asterisk indicates that all derivatives should be
taken at the symmetric fixed point $(A_{{E}},B_{{E}},
A_{{E}},A_{{E}},B_{{E}})$. The above equation gives, in
general, three eigenvalues for each $b$, but in practice it
turns out that only one of them (to be henceforth denoted
by $\lambda_{\phi}$) is relevant. Knowing $\lambda_{\phi}$
we can determine the critical exponent $\phi$ through the
formula
\begin{equation}\label{eq:fi}
\phi={\ln\lambda_\phi\over \ln\lambda_\nu}\>.
\end{equation}
Hence, in an exact RG evaluation of $\phi$ one needs to calculate
partial derivatives of sums (\ref{eq:RGA1})--(\ref{eq:RGB1}), and
thereby one should find the coefficients
$a_1(N_A,N_B,N_{A_1},N_{A_2},N_{B_1})$,
$a_2(N_A,N_B,N_{A_1},N_{A_2},N_{B_1})$, and
$b_1(N_A,N_B,N_{A_1},N_{A_2},N_{B_1})$
 by an exact enumeration of
all possible SAWs for each particular $b$,  which has been
accomplished for fractals with $b\le 4$. However, for $b\ge 5$, as
in the case of the bulk phase, the exact enumeration turns out to
be a formidable task. We have circumvent this problem by applying
the MCRG method. Within this method, the first step would be to
locate the symmetric fixed point. Fortunately, because of the
structure of the symmetric fixed, the results given in
Table~\ref{tabela1}
\Table{\label{tabela1} The MCRG ($2\le b\le
40$) results obtained in this work for the bulk fixed-point value
parameters $A_E$ and $B_E$, and the SAW critical exponents $\nu_E$
and $\phi_E$ for the 3d SG family of fractals. Each entry of the
table has been obtained by performing $10^5$ requisite Monte Carlo
simulation.}
\begin{tabular}{@{}llll} \br
$b$ & $(A_E,B_E)$ & $\nu_E$ & $\phi_E$
\\
\mr
  $\ $2$\ $ & $\ $(0.4311$\pm$0.0009, 0.0505$\pm$0.0023)$\ $&
  $\ $0.6742$\pm$0.0051$\ $ &$\ $0.7484$\pm$0.0152$\ $ \\
  3 & (0.3421$\pm$0.0004, 0.0245$\pm$0.0015) &0.6543$\pm$0.0021&
  0.7148$\pm$0.0037 \\
  4 & (0.2898$\pm$0.0004, 0.0122$\pm$0.0020) &0.6414$\pm$0.0012&
  0.6901$\pm$0.0036 \\
  5 & (0.2560$\pm$0.0004, 0.0067$\pm$0.0019)&0.6315$\pm$0.0010
  &0.6707$\pm$0.0028 \\
  6 & (0.2319$\pm$0.0003, 0.0038$\pm$0.0012) &0.6239$\pm$0.0009
  &0.6496$\pm$0.0024 \\
  7 & (0.2148$\pm$0.0003, 0.0020$\pm$0.0018) &0.6169$\pm$0.0006
  &0.6333$\pm$0.0019 \\
  8 & (0.2016$\pm$0.0003, 0.0012$\pm$0.0026)&0.6130$\pm$0.0005
  &0.6198$\pm$0.0016 \\
  9 & (0.1912$\pm$0.0004, 0.0007$\pm$0.0008) &0.6087$\pm$0.0006
  &0.6023$\pm$0.0017 \\
  10 & (0.1829$\pm$0.0003, 0.0005$\pm$0.0023) &0.6048$\pm$0.0003
  &0.5894$\pm$0.0013 \\
  12 & (0.1703$\pm$0.0004, 0.0001$\pm$0.0035) &0.5987$\pm$0.0003
  &0.5686$\pm$0.0012 \\
  15 & (0.1581$\pm$0.0001, -) &0.5933$\pm$0.0002
  &0.5385$\pm$0.0010 \\
  17 & (0.1526$\pm$0.0001, -) &0.5899$\pm$0.0002&0.5213$\pm$0.0010 \\
  20 & (0.1462$\pm$0.0001, -) &0.5869$\pm$0.0002&0.5014$\pm$0.0012 \\
  25 & (0.1399$\pm$0.0001, -) &0.5817$\pm$0.0002&0.4666$\pm$0.0008 \\
  30 & (0.1353$\pm$0.0001, -) &0.5804$\pm$0.0002&0.4325$\pm$0.0008 \\
  35 & (0.1327$\pm$0.0001, -) &0.5759$\pm$0.0001&0.4275$\pm$0.0007 \\
  40 & (0.1305$\pm$0.0001, -) &0.5755$\pm$0.0001&0.3765$\pm$0.0007 \\
 \br
\end{tabular}
\endTable provide the complete information for this fixed point
for a sequence of fractals with $2\le b\le 40$. The next step in
the MCRG method consists of finding $\lambda_\phi$, without
explicit calculation of the RG equation coefficients.

To solve the  eigenvalue problem (\ref{eq:det2}), so as to
learn $\lambda_\phi$, we need to find the requisite partial
derivatives. These derivatives can be related to various
averages of the numbers $N_{A_1}$, $N_{A_2}$, and
$N_{B_1}$,  of different SAW parts (of the types $A_1$,
$A_2$ and $B_1$) within a SAW path. Indeed, we may apply
the relation (\ref{eq:4}) where here $X$ and $Y$ stands for
any pair of quantities from the set $\{A_1, A_2, B_1\}$.
Therefore, to calculate the  derivatives (\ref{eq:4}) for
$X,Y\in\{A_1, A_2, B_1\}$ at the symmetric fixed point, one
needs the nine averages ($\langle N_{A_1}\rangle_{A'_1}^*$,
$\langle N_{A_2}\rangle_{A'_1}^*$, $\langle
N_{B_1}\rangle_{A'_1}^*$, $\langle
N_{A_1}\rangle_{A'_2}^*$, $\langle
N_{A_2}\rangle_{A'_2}^*$, $\langle
N_{B_1}\rangle_{A'_2}^*$, $\langle
N_{A_1}\rangle_{B'_1}^*$, $\langle
N_{A_2}\rangle_{B'_1}^*$, $\langle
N_{B_1}\rangle_{B'_1}^*$), which are all measurable through
MC simulations. Solving numerically the eigenvalue equation
(\ref{eq:det2}) we obtain $\lambda_\phi$, and, finally,
using relation ($\ref{eq:fi}$), we find the values of the
critical exponent $\phi$ (in the extended polymer phase),
which are presented in Table~\ref{tabela1}, and discussed
in the following paragraph.

\begin{figure}
\hskip1cm
\includegraphics[scale=0.4]{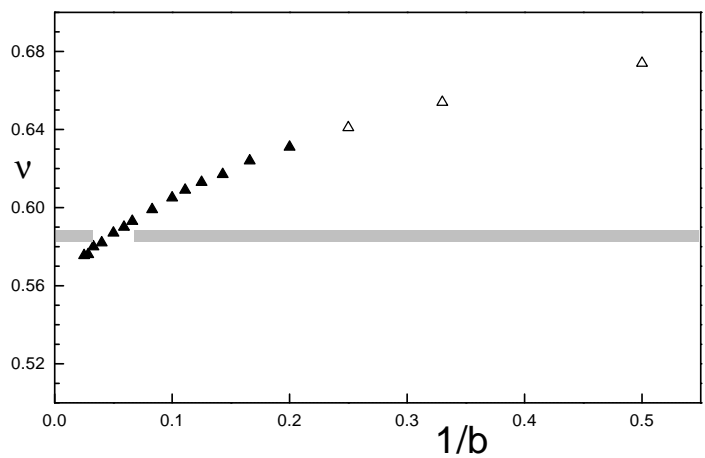} \caption{Data for the
end-to-end critical exponent $\nu$ for the 3d SG family of
fractals. The exact RG results are represented by open triangles,
while the MCRG results are depicted by solid triangles. The shaded
horizontal band represents the region of estimated values (found
in the literature) for the three--dimensional Euclidean critical
exponent $\nu$. The error bars related to the MCRG data are not
depicted in the figure since in all cases they lie within the
corresponding symbols.}
 \label{fig5}
 \end{figure}

\begin{figure}
 \hskip1cm
\includegraphics[scale=0.4]{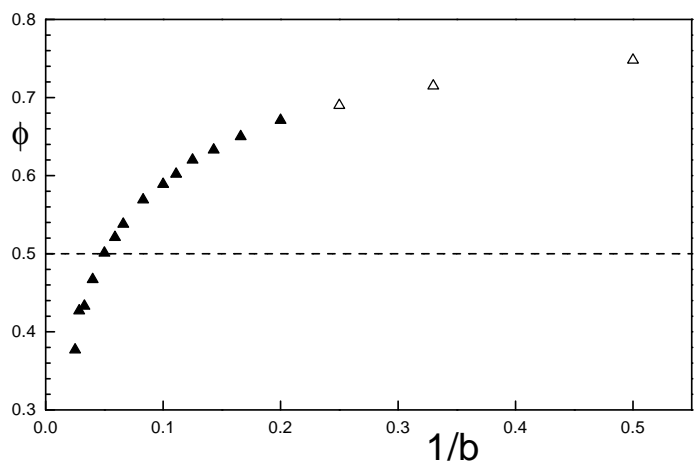}
\caption{Data for the adsorption critical exponent $\phi$ for the
3d SG family of fractals. The exact RG results are represented by
open triangles, while the MCRG results are depicted by solid
triangles. The dashed horizontal line represents the putative
universal Euclidean value 1/2 of crossover critical exponent
$\phi$. The error bars related to the MCRG data are not depicted
in the figure since in all cases they lie within the corresponding
symbols.}
 \label{fig6}
\end{figure}

First, we would like to compare the results obtained, via the
exact RG approach and through the MCRG technique, for the first
three members ($b=2,3,4$) of the SG fractal families (given in
Table~\ref{tabela-suki} and Table~\ref{tabela1}, respectively).
One can observe that the MCRG results for the critical exponents
$\nu$ and $\phi$ deviate at most 0.2\% from the exact results.
This very good agreement provides confidence in applying the MCRG
approach for a longer sequence of fractals ($5\le b\le40$).  For
the sake of a better assessment of the global behavior of the
critical exponents $\nu$ and $\phi$, as a function of the fractal
scaling parameter $b$, we depict the corresponding values (from
Tables~\ref{tabela-suki} and \ref{tabela1}) in Fig.~\ref{fig5}
 and
Fig.~\ref{fig6},
respectively.
 One can see that $\nu$, being a
monotonically decreasing function of $b$, crosses the narrow range
determined by the predictions made for the three--dimensional
Euclidean lattices, starting with $\nu=7/12=0.5833$ \cite{irene},
passing through the values $0.5850$ \cite{grassberger93} and
0.5874 \cite{prellberg01}, and ending with $\nu=0.5882$
\cite{nidras}. Crossing the interval of the possible Euclidean
values for the exponent $\nu$ occurs at the scaling fractal
parameter $b=20$. Interestingly, one may observe from
Fig.~\ref{fig6} that the exponent $\phi$, being monotonically
decreasing function of $b$, crosses the estimated
three--dimensional Euclidean value $\phi=1/2$ \cite{grassberger94}
also at $b=20$.

\section{Conclusion} \label{diskusija}

In this paper we studied the adsorption phenomenon of a
linear polymer, in a good and bad solvent, on impenetrable
boundaries of fractal containers modelled by the 3d SG
family of fractals. Each member of the 3d SG fractal family
has a fractal impenetrable 2d adsorbing boundary (which is,
in fact, 2d SG fractal surface) and can be labelled by an
integer $b$ ($2\le b\le\infty$). For the first three
members ($b=2,3,4$) of the 3d fractal family we have
performed exact RG analysis. This analysis enabled us to
establish the phase diagrams (for fractals with $b=3$ and
$b=4$; the $b=2$ case was studied previously
\cite{Bouchaud}), which turned out to be very rich from the
physical point of view. Indeed, the phase diagrams disclose
six different polymer phases that merge together at a
multi-critical point, whose nature for $b=3$ and $b=4$, as
well as the shape of the critical line that separates
adsorbed from desorbed phases, depend on the value of the
parameter $t$ (associated with the monomer energy in the
layer adjacent to the adsorbing boundary), which was not
the case for $b=2$. By analyzing the obtained phase
diagrams we may conclude that similar diagrams can be
expected for $b>4$, but finding their exact pictures
presently is not feasible.

By applying the exact and Monte Carlo renormalization group
(MCRG) method, we calculated the critical exponents $\nu$
(associated with the mean squared end-to-end distance of
polymers) and $\phi$ (associated with the number of
adsorbed monomers) for a sequence of fractals with $2\le
b\le4$ (exactly) and $2\le b\le40$ (Monte Carlo). The
reliability of the MCRG results is manifested by the fact
that in the cases $b=2,3,4$ the MCRG results for $\nu$ and
$\phi$ deviate at most 0.2\% from the exact results.
Unfortunately, it was possible to implement this powerful
(MCRG) method only in the case of the extended polymer
phase.

We find that, in the region studied, both $\nu$ and $\phi$
monotonically decrease with increasing $b$ (that is, with
increasing of the container fractal dimension $d_f$), and
the very interesting fact that both functions, $\nu(b)$ and
$\phi(b)$, cross the estimated Euclidean values at,
approximately, the same value of the scaling parameter
($b\approx20$). Here we would like to point out that in the
case of the 2d SG fractals both exponents $\nu$ and $\phi$
also cross the corresponding Euclidean values, but not at
the same value of $b$ (at $b\approx27$ for $\nu$
\cite{zivic1}, while at $b\approx6$ for $\phi$
\cite{zivic6}).

On the whole, our findings should be useful in making the
corresponding 3d models of the polymer adsorption phenomena in
porous media. Besides, our results may serve as beneficial in
constructing theories of the polymer adsorption phenomena for the
homogeneous 3d lattices, where so far, to the best of our
knowledge, an exact approach has not been made.

\ack{This paper has been done as a part of the work within the
project No.1634 funded by the Serbian Ministry of Science,
Technology and Development.}

\appendix

\section{\label{appendixa} RG equations for the surface parameters in the case
of the three-dimensional $b=3$ SG fractals}

In this Appendix we give the exact RG equations for the surface RG
parameters $A_1$, $A_2$ and $B_1$ in the case of the 3d $b=3$ SG
fractal. We have found that these equations have the following
form:
\begin{eqnarray}
A_1' &=&\sum_{i=0}^4F_{A_1}^i(A_1,A_2,B,B_1)A^i \, ,
\label{eq:A1}\\
A_2' &=&\sum_{i=0}^4F_{A_2}^i(A_1,A_2,B,B_1)A^i \, ,
\label{eq:A2}\\
B_1' &=&\sum_{i=0}^4F_{B_1}^i(A_1,A_2,B,B_1)A^i \, , \label{eq:B1}
\end{eqnarray}
where
\begin{eqnarray}
\fl F_{A_1}^0(A_1,A_2,B,B_1)=A_1^3+3A_1^4+A_1^5+2A_1^6\,
,\nonumber \\
\fl F_{A_2}^0(A_1,A_2,B,B_1)=B^4A_1A_2^2(44  A_2 + 132A_1 A_2 +
    132A_1 B_1 + 352 A_1^2  B_1)   \, ,\nonumber \\
\fl
F_{B_1}^0(A_1,A_2,B,B_1)=B^4(22A_2^3+132A_1^2A_2^2B_1+484A_1^3A_2B_1^2+484A_1^3B_1^3)\,\nonumber
, \end{eqnarray}
\begin{eqnarray}
\fl F_{A_1}^1= 2\,A_1\,A_2^2 +
  9\,{A_1}^2\,A_2^2 +
  10\,{A_1}^3\,A_2^2 +
  10\,{A_1}^4\,A_2^2 +
  8\,{A_1}^2\,A_2\,B_1 +16\,{A_1}^3\,A_2\,B_1 +\nonumber\\
\lo 20\,{A_1}^4\,A_2\,B_1 + 12\,{A_1}^2\,{B_1}^2 +
  6\,{A_1}^3\,{B_1}^2 +
  32\,{A_1}^4\,{B_1}^2+B^2( 6\,A_2^4 + \nonumber\\
\lo 14\,A_1\,A_2^4 +20\,{A_1}^2\,A_2^4 +64\,A_1\,A_2^3\,B_1 +
  32\,{A_1}^2\,A_2^3\,B_1 +
  40\,A_1\,A_2^2\,{B_1}^2 +\nonumber\\
\lo
  304\,{A_1}^2\,A_2^2\,{B_1}^2 +
  320\,{A_1}^2\,{B_1}^4)\, ,\nonumber\\
\fl F_{ A_1}^2= A_2^2 + 4\,A_1\,A_2^2 +
  9\,{A_1}^2\,A_2^2 +
  16\,{A_1}^3\,A_2^2 +
  14\,{A_1}^4\,A_2^2 + 3\,A_2^4 +
  9\,A_1\,A_2^4 +\nonumber\\
  \lo 12\,{A_1}^2\,A_2^4 +
8\,{A_1}^2\,A_2\,B_1 +
  36\,{A_1}^3\,A_2\,B_1 +
  40\,{A_1}^4\,A_2\,B_1 +4\,A_2^3\,B_1 +
   \nonumber\\
  \lo 24\,A_1\,A_2^3\,B_1 +
40\,{A_1}^2\,A_2^3\,B_1+
  12\,{A_1}^3\,{B_1}^2 +
  30\,{A_1}^4\,{B_1}^2 +
  \nonumber\\
  \lo 24\,A_1\,A_2^2\,{B_1}^2 +
  88\,{A_1}^2\,A_2^2\,{B_1}^2 +
  48\,A_1\,A_2\,{B_1}^3 +
  112\,{A_1}^2\,A_2\,{B_1}^3 +
  \nonumber\\
  \lo 128\,{A_1}^2\,{B_1}^4
+B( 4\,A_1\,A_2^2 +12\,{A_1}^2\,A_2^2 +
  8\,{A_1}^3\,A_2^2 +
16\,{A_1}^4\,A_2^2 + \nonumber\\
\lo 4\,A_2^4 +
  24\,A_1\,A_2^4 +
  40\,{A_1}^2\,A_2^4 +
  8\,{A_1}^2\,A_2\,B_1 +
  32\,{A_1}^3\,A_2\,B_1 +\nonumber\\
  \lo
  8\,{A_1}^4\,A_2\,B_1 +
  12\,A_2^3\,B_1 +
  48\,A_1\,A_2^3\,B_1 +
  120\,{A_1}^2\,A_2^3\,B_1 +24\,{A_1}^2\,{B_1}^2 +\nonumber\\
  \lo
64\,{A_1}^4\,{B_1}^2 +
  64\,A_1\,A_2^2\,{B_1}^2 +
  176\,{A_1}^2\,A_2^2\,{B_1}^2 +
  176\,A_1\,A_2\,{B_1}^3 +
\nonumber\\
  \lo  352\,{A_1}^2\,A_2\,{B_1}^3 +264\,{A_1}^2\,{B_1}^4)
+B^2( 6\,A_2^2 + 12\,{A_1}^2\,A_2^2 +
  \nonumber\\
  \lo 36\,{A_1}^3\,A_2^2 + 20\,A_2^4 +
  44\,A_1\,A_2^4 +24\,{A_1}^2\,A_2\,B_1 +
  24\,{A_1}^3\,A_2\,B_1 +\nonumber\\
  \lo
  96\,{A_1}^4\,A_2\,B_1 +
  64\,A_1\,A_2^3\,B_1 +232\,{A_1}^2\,A_2^3\,B_1 +36\,{A_1}^3\,{B_1}^2 +
  \nonumber\\
  \lo
  36\,{A_1}^4\,{B_1}^2 +
  168\,A_1\,A_2^2\,{B_1}^2 +
  168\,{A_1}^2\,A_2^2\,{B_1}^2 +
  928\,{A_1}^2\,A_2\,{B_1}^3)
\, ,\nonumber\\
\fl F_{A_1}^3=
2\,A_2^2+8\,A_1\,A_2^2+29\,{A_1}^2\,A_2^2+42\,{A_1}^3\,A_2^2+34\,{A_1}^4\,A_2^2
+ 4\,A_2^4 +10\,A_1\,A_2^4
+\nonumber\\
\lo
14\,{A_1}^2\,A_2^4+2\,A_2^6+28\,{A_1}^2\,A_2\,B_1+72\,{A_1}^3\,A_2\,B_1+
96\,{A_1}^4\,A_2\,B_1+\nonumber\\
\lo 4\,A_2^3\,B_1 +
24\,A_1\,A_2^3\,B_1+44\,{A_1}^2\,A_2^3\,B_1+4\,A_2^5\,B_1+24\,{A_1}^2\,{B_1}^2+
\nonumber\\
\lo 30\,{A_1}^3\,{B_1}^2+100\,{A_1}^4\,{B_1}^2+
26\,A_1\,A_2^2\,{B_1}^2+116\,{A_1}^2\,A_2^2\,{B_1}^2+\nonumber\\
\lo 16\,A_2^4\,{B_1}^2+
64\,A_1\,A_2\,{B_1}^3+112\,{A_1}^2\,A_2\,{B_1}^3+124\,{A_1}^2\,{B_1}^4+
\nonumber\\
\lo 136\,A_2^2\,{B_1}^4+B(4\,A_2^2+12\,A_1\,A_2^2+
28\,A_1^2\,A_2^2+40\,A_1^3\,A_2^2
+ \nonumber\\
\lo 40\,A_1^4\,A_2^2+ 24\,A_2^4 + 48\,A_1\,A_2^4 +
76\,A_1^2\,A_2^4
+ 20\,A_2^6 + 24\,A_1^2\,A_2\,B_1+ \nonumber\\
\lo 112\,A_1^3\,A_2\,B_1+88\,A_1^4\,A_2\,B_1+ 32\,A_2^3\,B_1 +
136\,A_1\,A_2^3\,B_1 + 192\,A_1^2\,A_2^3\,B_1
+ \nonumber\\
\lo 48\,A_2^5\,B_1+24\,A_1^2\,{B_1}^2+24\,A_1^3\,{B_1}^2+
136\,A_1^4\,{B_1}^2+96\,A_1\,A_2^2\,{B_1}^2+
\nonumber\\
\lo 768\,A_1^2\,A_2^2\,{B_1}^2+160\,A_2^4\,{B_1}^2 +
448\,A_1\,A_2\,{B_1}^3 +352\,A_1^2\,A_2\,{B_1}^3 +\nonumber\\
\lo  944\,A_1^2\,{B_1}^4 + 1320\,A_2^2\,{B_1}^4)\, ,\nonumber\\
\fl F_{A_1}^4=2\,A_2^2+8\,A_1\,A_2^2+26\,{A_1}^2\,A_2^2+
36\,{A_1}^3\,A_2^2+32\,{A_1}^4\,A_2^2+8\,A_2^4+26\,A_1\,A_2^4+
\nonumber\\
\lo
32\,{A_1}^2\,A_2^4+24\,{A_1}^2\,A_2\,B_1+72\,{A_1}^3\,A_2\,B_1+
80\,{A_1}^4\,A_2\,B_1+
\nonumber\\
\lo 8\,A_2^3\,B_1+64\,A_1\,A_2^3\,B_1+
116\,{A_1}^2\,A_2^3\,B_1+24\,{A_1}^2\,{B_1}^2+\nonumber\\
\lo 24\,{A_1}^3\,{B_1}^2+
100\,{A_1}^4\,{B_1}^2+74\,A_1\,A_2^2\,{B_1}^2+
\nonumber\\
\lo 236\,{A_1}^2\,A_2^2\,{B_1}^2+112\,A_1\,A_2\,{B_1}^3+
336\,{A_1}^2\,A_2\,{B_1}^3+292\,{A_1}^2\,{B_1}^4\, , \nonumber\\
\fl F_{A_2}^1=B(2\,A_2^3+4\,A_1\,A_2^3+12\,{A_1}^2\,A_2^3+
 6\,{A_1}^3\,A_2^3+6\,A_2^2\,B_1+12\,A_1\,A_2^2\,B_1 +\nonumber\\
\lo 32\,{A_1}^3\,A_2^2\,B_1+24\,{A_1}^2\,A_2\,{B_1}^2+
44\,{A_1}^3\,{B_1}^3)+B^3(24\,A_2^3+
  \nonumber\\
\lo 88\,A_1\,A_2^3+88\,{A_1}^2\,A_2^3+
168\,{A_1}^3\,A_2^3+32\,A_2^5+84\,A_1\,A_2^5+\nonumber\\
\lo 88\,A_1\,A_2^2\,B_1+
448\,{A_1}^2\,A_2^2\,B_1+88\,{A_1}^3\,A_2^2\,B_1+88\,A_2^4\,B_1+
  \nonumber\\
\lo 132\,A_1\,A_2^4\,B_1+640\,{A_1}^3\,A_2\,{B_1}^2+
920\,A_1\,A_2^3\,{B_1}^2+320\,A_2^2\,{B_1}^3)\, ,
\nonumber\\
\fl F_{A_2}^2=A_2+4\,A_1\,A_2+10\,{A_1}^2\,A_2+10\,{A_1}^3\,A_2+
14\,{A_1}^4\,A_2+8\,{A_1}^5\,A_2+2\,{A_1}^2\,B_1+\nonumber\\
\lo 8\,{A_1}^3\,B_1+4\,{A_1}^4\,B_1+18\,{A_1}^5\,B_1+
B(8\,A_2^3+24\,A_1\,A_2^3+ \nonumber\\
\lo 36\,{A_1}^2\,A_2^3+32\,{A_1}^3\,A_2^3
+12\,A_2^5+12\,A_1\,A_2^5+36\,A_1\,A_2^2\,B_1+\nonumber\\
\lo
76\,{A_1}^2\,A_2^2\,B_1+76\,{A_1}^3\,A_2^2\,B_1+8\,A_2^4\,B_1+68\,A_1\,A_2^4\,B_1+
64\,{A_1}^2\,A_2\,{B_1}^2+\nonumber\\
\lo
88\,{A_1}^3\,A_2\,{B_1}^2+48\,A_2^3\,{B_1}^2+88\,{A_1}^3\,{B_1}^3+
304\,A_1\,A_2^2\,{B_1}^3)+B^2(6\,A_2+\nonumber\\
\lo 12\,{A_1}^2\,A_2+36\,{A_1}^3\,A_2+ 12\,{A_1}^4\,A_2
+24\,{A_1}^5\,A_2 + 36\,A_2^3
+56\,A_1\,A_2^3+\nonumber\\
\lo 104\,{A_1}^2\,A_2^3+96\,{A_1}^3\,A_2^3+
50\,A_2^5+68\,A_1\,A_2^5+12\,{A_1}^2\,B_1+ \nonumber\\
\lo 24\,{A_1}^4\,B_1+ 12\,{A_1}^5\,B_1+12\,A_2^2\,B_1+
88\,A_1\,A_2^2\,B_1+160\,{A_1}^2\,A_2^2\,B_1+ \nonumber\\
\lo 296\,{A_1}^3\,A_2^2\,B_1+72\,A_2^4\,B_1+ 252\,A_1\,A_2^4\,B_1+
168\,A_1\,A_2\,{B_1}^2+\nonumber\\
\lo
216\,{A_1}^2\,A_2\,{B_1}^2+168\,{A_1}^3\,A_2\,{B_1}^2+160\,A_2^3\,{B_1}^2
+
328\,A_1\,A_2^3\,{B_1}^2+\nonumber\\
\lo 328\,{A_1}^3\,{B_1}^3+132\,A_2^2\,{B_1}^3+952\,A_1\,A_2^2\,{B_1}^3)\nonumber\\
\fl F_{A_2}^3=2\,A_2+4\,A_1\,A_2+12\,{A_1}^2\,A_2 +
16\,{A_1}^3\,A_2+16\,{A_1}^4\,A_2 +12\,{A_1}^5\,A_2 + 8\,A_2^3 +\nonumber\\
\lo 24\,A_1\,A_2^3+42\,{A_1}^2\,A_2^3+26\,{A_1}^3\,A_2^3+
4\,{A_1}^2\,B_1+8\,{A_1}^3\,B_1+8\,{A_1}^4\,B_1+\nonumber\\
\lo 20\,{A_1}^5\,B_1+4\,A_2^2\,B_1+36\,A_1\,A_2^2\,B_1+
68\,{A_1}^2\,A_2^2\,B_1+92\,{A_1}^3\,A_2^2\,B_1+ \nonumber\\
\lo 20\,A_1\,A_2\,{B_1}^2+ 76\,{A_1}^2\,A_2\,{B_1}^2+
112\,{A_1}^3\,A_2\,{B_1}^2+120\,{A_1}^3\,{B_1}^3+\nonumber\\
\lo B(4\,A_2+8\,A_1\,A_2+  24\,{A_1}^2\,A_2+
32\,{A_1}^3\,A_2+32\,{A_1}^4\,A_2+\nonumber\\
\lo
24\,{A_1}^5\,A_2+22\,A_2^3+64\,A_1\,A_2^3+100\,{A_1}^2\,A_2^3+
86\,{A_1}^3\,A_2^3+14\,A_2^5+\nonumber\\
\lo 32\,A_1\,A_2^5+8\,{A_1}^2\,B_1+
16\,{A_1}^3\,B_1+16\,{A_1}^4\,B_1+40\,{A_1}^5\,B_1+\nonumber\\
\lo 6\,A_2^2\,B_1 +60\,A_1\,A_2^2\,B_1+236\,{A_1}^2\,A_2^2\,B_1+
208\,{A_1}^3\,A_2^2\,B_1+28\,A_2^4\,B_1+\nonumber\\
\lo 28\,A_1\,A_2^4\,B_1 +64\,A_1\,A_2\,{B_1}^2+
88\,{A_1}^2\,A_2\,{B_1}^2+304\,{A_1}^3\,A_2\,{B_1}^2 +\nonumber\\
\lo
320\,A_1\,A_2^3\,{B_1}^2+132\,{A_1}^3\,{B_1}^3+132\,A_2^2\,{B_1}^3)\,
,\nonumber\\
\fl F_{A_2}^4=
2\,A_2+4\,A_1\,A_2+12\,{A_1}^2\,A_2+16\,{A_1}^3\,A_2 +
16\,{A_1}^4\,A_2+12\,{A_1}^5\,A_2 +10\,A_2^3 + \nonumber\\
\lo 30\,A_1\,A_2^3+50\,{A_1}^2\,A_2^3+40\,{A_1}^3\,A_2^3+
8\,A_2^5+12\,A_1\,A_2^5 +4\,{A_1}^2\,B_1+\nonumber\\
\lo 8\,{A_1}^3\,B_1 +
8\,{A_1}^4\,B_1+20\,{A_1}^5\,B_1+6\,A_2^2\,B_1+
40\,A_1\,A_2^2\,B_1+\nonumber\\
\lo
94\,{A_1}^2\,A_2^2\,B_1+112\,{A_1}^3\,A_2^2\,B_1+20\,A_2^4\,B_1+58\,A_1\,A_2^4\,B_1+\nonumber\\
\lo 20\,A_1\,A_2\,{B_1}^2+
68\,{A_1}^2\,A_2\,{B_1}^2+108\,{A_1}^3\,A_2\,{B_1}^2+36\,A_2^3\,{B_1}^2+\nonumber\\
\lo 92\,A_1\,A_2^3\,{B_1}^2+100\,{A_1}^3\,{B_1}^3+
32\,A_2^2\,{B_1}^3+192\,A_1\,A_2^2\,{B_1}^3\, ,\nonumber\\
\fl
F_{B_1}^1=B(3\,{A_1}^2\,A_2^3+8\,{A_1}^3\,A_2^3+6\,{A_1}^3\,A_2^2\,B_1+
 44\,{A_1}^3\,A_2\,{B_1}^2 +66\,{A_1}^2\,{B_1}^3)+\nonumber\\
\lo B^3( 44\,A_1\,A_2^3 +44\,{A_1}^2\,A_2^3+22\,A_2^5+
320\,{A_1}^3\,A_2^2\,B_1 +66\,A_2^4\,B_1 +\nonumber\\
\lo 120\,A_1\,A_2^4\,B_1
+312\,{A_1}^2\,A_2\,{B_1}^2+312\,{A_1}^3\,A_2\,{B_1}^2+
160\,A_1\,A_2^3\,{B_1}^2 +\nonumber\\
\lo 968\,{A_1}^3\,{B_1}^3 + 968\,A_1\,A_2^2\,{B_1}^3+4308\,A_1\,{B_1}^5)\, ,\nonumber\\
\fl F_{B_1}^2={A_1}^3\,A_2+7\,{A_1}^4\,A_2+8\,{A_1}^5\,A_2+
15\,{A_1}^4\,B_1+8\,{A_1}^5\,B_1+B(6\,A_1\,A_2^3+\nonumber\\
\lo
12\,{A_1}^2\,A_2^3+12\,{A_1}^3\,A_2^3+10\,A_1\,A_2^5+16\,{A_1}^2\,A_2^2\,B_1+
60\,{A_1}^3\,A_2^2\,B_1+\nonumber\\
\lo 24\,A_1\,A_2^4\,B_1
+44\,{A_1}^2\,A_2\,{B_1}^2+176\,{A_1}^3\,A_2\,{B_1}^2+
88\,A_1\,A_2^3\,{B_1}^2 +  \nonumber\\
\lo 132\,{A_1}^3\,{B_1}^3
+576\,A_1\,A_2\,{B_1}^4)+B^2(6\,{A_1}^3\,A_2+
\,{A_1}^4\,A_2+ \nonumber\\
\lo 54\,{A_1}^2\,A_2^3+48\,{A_1}^3\,A_2^3
+6\,A_2^5+28\,A_1\,A_2^5+18\,{A_1}^4\,B_1+\nonumber\\
\lo 48\,{A_1}^5\,B_1+
128\,{A_1}^2\,A_2^2\,B_1+136\,{A_1}^3\,A_2^2\,B_1 +18\,A_2^4\,B_1+\nonumber\\
\lo 124\,A_1\,A_2^4\,B_1+432\,{A_1}^3\,A_2\,{B_1}^2+
318\,A_1\,A_2^3\,{B_1}^2+132\,{A_1}^2\,{B_1}^3 +\nonumber\\
\lo 632\,{A_1}^3\,{B_1}^3+312\,A_1\,A_2^2\,{B_1}^3+
1936\,A_1\,A_2\,{B_1}^4+1452\,A_1\,{B_1}^5)\, ,
\nonumber\\
\fl F_{B_1}^3=2\,{A_1}^3\,A_2+10\,{A_1}^4\,A_2+8\,{A_1}^5\,A_2+
2\,A_1\,A_2^3+14\,{A_1}^2\,A_2^3+22\,{A_1}^3\,A_2^3
+\nonumber\\
\lo 18\,{A_1}^4\,B_1+16\,{A_1}^5\,B_1+26\,{A_1}^2\,A_2^2\,B_1+
74\,{A_1}^3\,A_2^2\,B_1+\nonumber\\
\lo
12\,{A_1}^2\,A_2\,{B_1}^2+164\,{A_1}^3\,A_2\,{B_1}^2+44\,{A_1}^2\,{B_1}^3+
120\,{A_1}^3\,{B_1}^3+\nonumber\\
\lo B(4\,{A_1}^3\,A_2+20\,{A_1}^4\,A_2+16\,{A_1}^5\,A_2+4\,A_2^3+12\,A_1\,A_2^3+\nonumber\\
\lo
33\,{A_1}^2\,A_2^3+24\,{A_1}^3\,A_2^3+8\,A_2^5+36\,{A_1}^4\,B_1
+32\,{A_1}^5\,B_1+\nonumber\\
\lo 72\,{A_1}^2\,A_2^2\,B_1+174\,{A_1}^3\,A_2^2\,B_1+
18\,A_2^4\,B_1+48\,A_1\,A_2^4\,B_1+\nonumber\\
\lo 88\,{A_1}^2\,A_2\,{B_1}^2+352\,{A_1}^3\,A_2\,{B_1}^2+
66\,A_1\,A_2^3\,{B_1}^2+66\,{A_1}^2\,{B_1}^3+\nonumber\\
\lo
592\,{A_1}^3\,{B_1}^3+312\,A_1\,A_2^2\,{B_1}^3+1452\,A_1\,{B_1}^5)\,
, \nonumber\\
\fl F_{B_1}^4=2\,{A_1}^3\,A_2+10\,{A_1}^4\,A_2+8\,{A_1}^5\,A_2+
A_2^3+6\,A_1\,A_2^3+17\,{A_1}^2\,A_2^3+\nonumber\\
\lo 16\,{A_1}^3\,A_2^3
+3\,A_2^5+8\,A_1\,A_2^5+18\,{A_1}^4\,B_1+16\,{A_1}^5\,B_1+\nonumber\\
\lo
26\,{A_1}^2\,A_2^2\,B_1+72\,{A_1}^3\,A_2^2\,B_1+7\,A_2^4\,B_1+32\,A_1\,A_2^4\,B_1+\nonumber\\
\lo 44\,{A_1}^2\,A_2\,{B_1}^2+
202\,{A_1}^3\,A_2\,{B_1}^2+64\,A_1\,A_2^3\,{B_1}^2+\nonumber\\
\lo 66\,{A_1}^2\,{B_1}^3+230\,{A_1}^3\,{B_1}^3+
100\,A_1\,A_2^2\,{B_1}^3+\nonumber\\
\lo 440\,A_1\,A_2\,{B_1}^4+472\,A_1\,{B_1}^5\,.\nonumber
\end{eqnarray}

\section{RG equations for the bulk parameters in the case of the
three-dimensional $b=4$ SG fractals}

The exact recursion relations for the bulk RG parameters $A$ and
$B$ in the case of the 3d $b=4$ Sierpinski fractal have the
following form
\begin{eqnarray}
\fl A'= {A^4} + 12\,{A^5} + 62\,{A^6} + 220\,{A^7} + 782\,{A^8} +
2426\,{A^9} +12\,{A^5}\,B + 128\,{A^6}\,B + \nonumber\\
\lo 776\,{A^7}\,B +  3416\,{A^8}\,B + 13324\,{A^9}\,B
+18\,{A^4}\,{B^2} + 72\,{A^5}\,{B^2} +  432\,{A^6}\,{B^2} +
 \nonumber\\
\lo 2736\,{A^7}\,{B^2}+ 13294\,{A^8}\,{B^2} +56004\,{A^9}\,{B^2} +
  19\,{A^5}\,{B^3} + 1456\,{A^6}\,{B^3} +  \nonumber\\
 \lo 8704\,{A^7}\,{B^3}
  +48256\,{A^8}\,{B^3} +
   213968\,{A^9}\,{B^3} + 36\,{A^4}\,{B^4} +
  688\,{A^5}\,{B^4} + \nonumber\\
 \lo  6880\,{A^6}\,{B^4} +33264\,{A^7}\,{B^4} +
  173936\,{A^8}\,{B^4} +
  816\,{A^5}\,{B^5} + 781456\,{A^7}\,{B^7} +\nonumber\\
  \lo
  4350864\,{A^8}\,{B^7} + 2904\,{A^4}\,{B^8} +31616\,{A^5}\,{B^8} +
  248608\,{A^6}\,{B^8} +\nonumber\\
  \lo
2047360\,{A^7}\,{B^8} +
  9995376\,{A^8}\,{B^8} + 70400\,{A^5}\,{B^9} + 779584\,{A^6}\,{B^9} +\nonumber\\\lo
  5688896\,{A^7}\,{B^9} +13632\,{A^6}\,{B^5} +10369\,{A^7}\,{B^5} +
   562192\,{A^8}\,{B^5} +\nonumber\\
   \lo
   640\,{A^4}\,{B^6}
  +  1792\,{A^5}\,{B^6}
  32432\,{A^6}\,{B^6} + 250672\,{A^7}\,{B^6} +1552992\,{A^8}\,{B^6} +\nonumber\\
  \lo
   7712\,{A^5}\,{B^7} + 98528\,{A^6}\,{B^7}+   24273344\,{A^8}\,{B^9} +
    65568\,{A^4}\,{B^{10}} +\nonumber\\
  \lo  354336\,{A^5}\,{B^{10}} + 2960032\,{A^6}\,{B^{10}}+
  11284384\,{A^7}\,{B^{10}} + 44440800\,{A^8}\,{B^{10}} + \nonumber\\
  \lo  139392\,{A^5}\,{B^{11}} + 6933440\,{A^6}\,{B^{11}} +
   25364736\,{A^7}\,{B^{11}} +20945408\,{A^8}\,{B^{11}} +\nonumber\\
   \lo
  259424\,{A^4}\,{B^{12}}+
   3408768\,{A^5}\,{B^{12}} + 7491712\,{A^6}\,{B^{
12}} \, ,\nonumber\\
\fl B'= A^8 + 24\,A^9 + 268\,A^{10} + 1624\,A^{11} +6544\,A^{12}
+20288\,A^{13} +  50676\,A^{14} + \nonumber\\
\lo 103904\,A^{15} +173050\,A^{16} + 225108\,A^{17}+215392\,A^{18}
+
134968\,A^{19} + \nonumber\\
\lo  42514\,A^{20}+  328\,A^9\,B +2584\,A^{10}\,B +
13940\,A^{11}\,B
 + 56768\,A^{12}\,B +  \nonumber\\
\lo  188356\,A^{13}\,B +491496\,A^{14}\,B + 1000000\,A^{15}\,B
+1539632\,A^{16}\,B + \nonumber\\
\lo 1701920\,A^{17}\,B
+1192152\,A^{18}\,B +
  397392\,A^{19}\,B + 36\,A^8\,B^2 + \nonumber\\
\lo  832\,A^9\,B^2 +
  14176\,A^{10}\,B^2 +85720\,A^{11}\,B^2+368404\,A^{12}\,B^2 +
  \nonumber\\
\lo 1199040\,A^{13}\,B^2+  2954904\,A^{14}\,B^2 +
5370136\,A^{15}\,B^2 + 6814424\,A^{16}\,B^2 +\nonumber\\
\lo
   5299288\,A^{17}\,B^2 + 1901008\,A^{18}\,B^2 + 2112\,A^8\,B^3 +  17424\,A^9\,B^3 +\nonumber\\
\lo    89344\,A^{10}\,B^3 +
  482272\,A^{11}\,B^3 +1965376\,A^{12}\,B^3 +5900928\,A^{13}\,B^3 +\nonumber\\
\lo     12721232\,A^{14}\,B^3 +
   18817920\,A^{15}\,B^3 + 16698368\,A^{16}\,B^3 +\nonumber\\
\lo 6662824\,A^{17}\,B^3 +616\,A^6\,B^4+1584\,A^7\,B^4 +
  6388\,A^8\,B^4 + \nonumber\\
\lo 76968\,A^9\,B^4+523696\,A^{10}\,B^4 + 2577592\,A^{11}\,B^4 +
  9113852\,A^{12}\,B^4 +  \nonumber\\
\lo 23214816\,A^{13}\,B^4 + 40127080\,A^{14}\,B^4 +
41398192\,A^{15}\,B^4 +\nonumber\\
\lo
  18548660\,A^{16}\,B^4 + 12912\,A^7\,B^5+62560\,A^8\,B^5+  451600\,A^9\,B^5 + \nonumber\\
\lo  2759488\,A^{10}\,B^5 +
  11688304\,A^{11}\,B^5 +34977824\,A^{12}\,B^5 +\nonumber\\
\lo   70124768\,A^{13}\,B^5 +
    82557440\,A^{14}\,B^5 +41200784\,A^{15}\,B^5 + \nonumber\\
\lo  1584\,A^6\,B^6 + 45056\,A^7\,B^6 +  296744\,A^8\,B^6 +
  2361856\,A^9\,B^6 +\nonumber\\
\lo  12623808\,A^{10}\,B^6 +  43429664\,A^{11}\,B^6 +
  102376032\,A^{12}\,B^6 + \nonumber\\
\lo  139837056\,A^{13}\,B^6 +78454776\,A^{14}\,B^6 +
7744\,A^6\,B^7 + 133440\,A^7\,B^7 +
  \nonumber\\
\lo  1858832\,A^8\,B^7 +  10467744\,A^9\,B^7 +
46337232\,A^{10}\,B^7 +\nonumber\\
\lo 126642624\,A^{11}\,B^7 +
   201751952\,A^{12}\,B^7 + 126633376\,A^{13}\,B^7 +   \nonumber\\
\lo  17232\,A^5\,B^8 + 54192\,A^6\,B^8 +
  950784\,A^7\,B^8 + 8336624\,A^8\,B^8 +\nonumber\\
\lo  42390624\,A^9\,B^8 +139673656\,A^{10}\,B^8 +
253941232\,A^{11}\,B^8 +
   \nonumber\\
\lo 172479256\,A^{12}\,B^8 +  431776\,A^6\,B^9 + 6405824\,A^7\,B^9
+32770448\,A^8\,B^9 +\nonumber\\
\lo
  127799456\,A^9\,B^9 +
   273263456\,A^{10}\,B^9 +  204194352\,A^{11}\,B^9 + \nonumber\\
\lo 471328\,A^5\,B^{10} +
    4395872\,A^6\,B^{10} +21993584\,A^7\,B^{10} + \nonumber\\
    \lo 107371568\,A^8\,B^{10} +
  262855104\,A^9\,B^{10} +
  199987864\,A^{10}\,B^{10} +\nonumber\\
 \lo   1224960\,A^5\,B^{11} + 17655616\,A^6\,B^{11} +76932128\,A^7\,B^{11} +
   \nonumber\\
   \lo 220464528\,A^8\,B^{11} +
  151443088\,A^9\,B^{11} + 3702272\,A^4\,B^{12} + \nonumber\\
  \lo 11642752\,A^5\,B^{12} +
  66511552\,A^6\,B^{12} +148524304\,A^7\,B^{12} + \nonumber\\
  \lo
  99652120\,A^8\,B^{12} + 3748096\,A^3\,B^{13} +
  30934336\,A^5\,B^{13} +\nonumber\\
  \lo   110065872\,A^6\,B^{13} + 26465392\,A^7\,B^{13} +
  6830208\,A^3\,B^{14} +\nonumber\\
  \lo 30509600\,A^4\,B^{14} +
   49942336\,A^5\,B^{14} +  20614528\,A^3\,B^{15} + 5153632\,B^{16}\, .\nonumber
 \end{eqnarray}


\section*{References}

\begin{thebibliography}{10}




\bibitem{napper}For review see  Jones R A L and
Richards R W 1999 {\it Polymers at Surfaces and Interfaces} (New
York: Cambridge University Press); Kawaguchi M and Takahashi A
1992 {\it Adv. Colloid Interface Sci.} {\bf 37} 219


\bibitem{binder} Eisenriegler E 1993 {\it Polymers near
Surfaces} (Singapore: World Scientific)

\bibitem{Dhar2001} Rajesh R, Dhar D, Giri D, Kumar S and Singh Y 2002
\PR {\em E} in print

\bibitem{Vrbova}  Vrbov\'a T and Whittington S G 1998  \JPA {\bf 31}
3989

\bibitem{Bouchaud} Bouchaud E and Vannimenus J 1989  \JP {\bf
50} 2931

\bibitem{Knezevic} Kne\v zevi\' c M and Vannimenus J 1987 \JPA {\bf
20} L969

\bibitem{DharVannimenus} Dhar D and Vannimenus J 1987 \JPA {\bf 20}
199

\bibitem{Jelena} Mari\v ci\' c J and Elezovi\' c--Had\v zi\' c S
 {\em Preprint} cond-mat/0205509

\bibitem{EKM} Elezovi\' c S, Kne\v zevi\' c M  and Milo\v sevi\' c S \JPA 1987 {\bf 20} 1215

\bibitem{Redner} Redner S and Reynolds P J 1981 \JPA
 {\bf 14} 2679


\bibitem{zivic1} Milo\v sevi\'c  S and   \v Zivi\'c I 1991 \JPA
  {\bf 24} L833

\bibitem{zivic5} Milo\v sevi\'c  S and   \v Zivi\'c I 1993 \JPA
  {\bf 26} 7263

\bibitem{irene} Hueter I {\em Preprint} math.PR/0108199

\bibitem{grassberger93}Grassberger P 1993 \JPA
  {\bf 26} 2769

\bibitem{prellberg01} Prellberg T 2001 \JPA {\bf 34} L599

\bibitem{nidras} Nidras P 1996 \JPA {\bf 29} 7929

\bibitem{grassberger94} Grassberger P 1994
 \JPA {\bf 27} 4069

\bibitem{zivic6}\v Zivi\'c I, Milo\v sevi\'c S and
Stanley H E 1994 {\it Phys. Rev. E} {\bf 49} 636


\end{thebibliography}
\end{document}